\documentclass[11pt]{article}
\usepackage[round,authoryear]{natbib}
\usepackage{authblk}
\usepackage{float}
\usepackage{sectsty}
\usepackage{graphicx}
\usepackage{caption}
\usepackage{subcaption}
\usepackage{url} 
\usepackage{multirow}
\usepackage{blindtext}
\usepackage{titlesec}
\usepackage{setspace}   
\usepackage{threeparttable}
\usepackage{amsmath}
\providecommand{\keywords}[1]
{
  \small	
  \textbf{\textit{Keywords---}} #1
}

\title{AI Feedback Enhances Community-Based Content Moderation through Engagement with Counterarguments}
\author[1, 2]{Saeedeh Mohammadi}
\author[1, 2, 3,*]{Taha Yasseri}
\affil[1]{School of Mathematics and Statistics,
  University College Dublin, Dublin, Ireland}
\affil[2]{School of Social Sciences and Philosophy,
  Trinity College Dublin, Dublin, Ireland}
\affil[3]{Faculty of Arts and Humanities,
  Technological University Dublin, Dublin,
Ireland}
  \affil[*]{The corresponding author}
\date{}

\begin{document}
\maketitle	
\doublespacing

\begin{abstract}
  Today, social media platforms are significant sources of news and political communication, but their role in spreading misinformation has raised significant concerns. In response, these platforms have implemented various content moderation strategies. One such method, Community Notes (formerly Birdwatch) on X (formerly Twitter), relies on crowdsourced fact-checking and has gained traction. However, it faces challenges such as partisan bias and delays in verification. This study explores an AI-assisted hybrid moderation framework in which participants receive AI-generated feedback, supportive, neutral, or argumentative, on their notes and are asked to revise them accordingly. The results show that incorporating feedback improves note quality, with the most substantial gains coming from argumentative feedback. This underscores the value of diverse perspectives and direct engagement in human-AI collective intelligence. The research contributes to ongoing discussions about AI’s role in political content moderation, highlighting the potential of generative AI and the importance of informed design.
\end{abstract}
\keywords{Content Moderation; Human-AI Interaction; Large Language Models; Polarisation; Hybrid Intelligence}

\pagebreak

\section{Introduction}
The emergence of social media has transformed how users consume and share information. Studies show that these platforms have become primary news sources for many individuals \citep{bakshy2015exposure, sunstein2001echo} and are widely used by politicians as campaign tools \citep{king2017news}. This shift presents challenges for platform owners in monitoring and verifying the information being shared. The widespread dissemination of misleading content and misinformation has become a primary concern for social media. This issue is especially pronounced during significant socio-political events, such as the 2016 U.S. presidential election \citep{allcott2017social} and the Brexit referendum \citep{greene2021misremembering}. The spread of false information during times of crisis can seriously mislead the public. For example, the rapid circulation of misinformation about COVID-19 led the WHO's Director-General to describe the situation as an ``infodemic'' \citep{who_munich_security_conference}. Another striking case is the January 6th, 2021, attack on the U.S. Capitol, where social media posts helped spread the misinformation that incited violence \citep{klein2025mobilising}. In response to these challenges, social media platforms have implemented various moderation strategies, including both community-based and automated approaches.

On January 7th, 2025, Mark Zuckerberg announced the end of Meta Platforms’ third-party fact-checking program, replacing it with Community Notes (formerly Birdwatch) \citep{meta2025}. Initially launched on X (formerly Twitter) in January 2021, Community Notes represents the first large-scale, community-based initiative to moderate misleading content \citep{communityNotes}. Community Notes relies on the wisdom of the crowd, leveraging the diversity of contributors to produce moderated outcomes. Contributors can write notes on potentially misleading posts to provide additional context, while other contributors evaluate these notes by rating their helpfulness. Notes that receive positive ratings from users with diverse perspectives are then displayed underneath the post in users' timelines \citep{communityNotes}.

Despite its promise, evidence from the first four years of Community Notes shows that only 13.55\% of posts with at least one proposed note ultimately have a note displayed on the timeline, and even among those, the process takes an average of 26 hours \citep{mohammadi2025birdwatch}. This delay undermines the effectiveness of the content moderation process \citep{truong2025delayed}, rendering it ineffective for addressing the majority of potentially harmful content.

These limitations highlight a broader, unresolved challenge in content moderation. While community-based systems such as Community Notes preserve human judgment and leverage ideological diversity, they remain constrained by partisan dynamics \citep{allen2022birds, yasseri2023can} and slow consensus-building processes \citep{mohammadi2025birdwatch}. In contrast, automated and LLM-based approaches offer speed and scalability, but raise concerns about bias, hallucinations, and user overreliance on machine-generated outputs.

Despite substantial progress in both paradigms, existing research has largely examined them in isolation, leaving open the question of how to effectively integrate their complementary strengths. In particular, there is limited understanding of how AI can be designed not merely as a substitute for human judgment, but as a collaborative partner that enhances human reasoning and facilitates more effective coordination. Addressing this gap is critical for improving the quality and responsiveness of community-based content moderation systems.

\subsection{Existing Literature}

The shift in content moderation strategies, from expert-driven and automated fact-checking to crowd-based approaches, reflects the limitations of earlier methods. Expert-driven moderation relies on human experts to review and classify large volumes of content, making it both costly and difficult to scale \citep{hassan2015quest}. Moreover, such approaches have faced criticism for potential political and cultural bias. For instance, a 2020 report found that 90\% of Republicans in the United States believed that social media platforms censor content based on political views \citep{americans2020Vogels}.

To address challenges related to scale and speed, platforms turned to automated moderation systems. Early efforts in this area focused on applying machine learning techniques to assess the truthfulness of statements \citep{demartini2020human}. These systems typically rely on supervised learning and require pre-labelled datasets, such as LIAR \citep{wang2017liar}, which compiles fact-checked claims from dedicated websites, and FEVER \citep{thorne2018fever}, which includes claims derived from Wikipedia. However, developing high-performing models remains challenging due to the need for large and diverse labelled datasets—a process that requires substantial labour from experts or crowdworkers \citep{vidgen2020directions}. Furthermore, the effectiveness of these systems is constrained by the quality and biases of their training data, which can perpetuate classification inaccuracies \citep{binns2017like}.

More recent approaches leverage larger benchmark datasets \citep{augenstein2019multifc, shaar2020known} and advanced methodologies, including deep learning \citep{kaliyar2021fakebert} and generative AI \citep{tang2024minicheck, ma2025local}. A growing body of work demonstrates that Large Language Models (LLMs) can effectively assess the credibility of news sources \citep{yang2025accuracy} and verify factual claims \citep{quelle2024perils, hoes2023leveraging, kuznetsova2025generative}, with performance improving when models are fine-tuned on augmented datasets \citep{zhou2024correcting}. 

Despite these advancements, such approaches remain constrained by their reliance on training data. LLM-based systems are also susceptible to hallucinations \citep{ma2025local} and often struggle to incorporate social and contextual signals. In addition, the lack of transparency in these systems can contribute to public scepticism toward their use \citep{agunlejika2025ai}. Moreover, purely automated approaches tend to perform less effectively in the presence of social influence \citep{lu2022effects}. Compounding these concerns, users frequently place considerable trust in LLM outputs—even when they are incorrect—raising the risk of overreliance on AI-generated content \citep{nguyen2018believe}.

Community-based moderation approaches have emerged as a promising alternative to both expert-driven and fully automated systems. By leveraging a large and diverse pool of contributors, these approaches are better positioned to address the scale of content moderation challenges. Unlike automated systems, they are not constrained by potentially biased training datasets and can incorporate nuanced social and contextual signals when evaluating content. Furthermore, because moderation decisions are collectively produced and publicly visible, community-based systems may help alleviate concerns about political bias and censorship. Empirical evidence supports these advantages: initial studies show that Community Notes can be more effective than experts at identifying misleading content, with substantial agreement between contributors and professional fact-checkers \citep{saeed2022crowdsourced}, and are more scalable and cost-efficient at the same time \citep{martel2024crowds}.

Nevertheless, Community Notes in its current form remains limited. Early evidence points to partisan behaviour among contributors. \citet{yasseri2023can} found that the network connecting note writers and raters mirrors the highly polarised friendship network on X. Similarly, \citet{allen2022birds} showed that contributors are more likely to write notes on posts by counter-partisans, label them as misleading, and rate notes from counter-partisans as unhelpful, while rating notes from co-partisans as helpful.

To mitigate these biases, Community Notes adjusted its rating algorithm. Notes that receive helpful ratings from contributors with diverse perspectives, who typically rate content differently, are now prioritised over those endorsed only by ideologically similar contributors \citep{wojcik2022birdwatch}. While this design aims to reduce partisan influence, it introduces a trade-off: it substantially slows down the process by which notes are selected and displayed. A recent report found that 74\% of accurate Community Notes addressing misinformation about the 2024 U.S. presidential election never reached users, allowing misleading content to go unchallenged. As a result, misinformation posts about the election spread 13 times faster than those accompanied by fact-checked notes \citep{CCDH2024}. These delays undermine the moderation process, as research shows that lag in content moderation significantly reduces its effectiveness \citep{truong2025delayed}.

A growing body of research has explored ways to improve both the quality and efficiency of Community Notes. Research suggests that contributors' political diversity can be leveraged at the note-writing stage, rather than solely at the rating stage \citep{yasseri2023can}. In a recent study, researchers paired participants to collaboratively write Community Notes and found that duos produced more helpful notes than individuals, with politically diverse pairs being particularly effective when evaluating Republican-leaning content \citep{juncosa2026benefit}.

This approach improves note quality while introducing additional coordination costs \citep{straub2023cost}. Requiring contributors with differing ideologies to be available simultaneously and collaborate effectively may increase the time needed to produce a note, potentially exacerbating existing delays rather than alleviating them.

Another prominent direction involves integrating large language models (LLMs) into the note generation process. For instance, \citet{de2025supernotes} fine-tuned an LLM on the Community Notes dataset to generate “supernotes”, which were rated as significantly more helpful than notes written by human contributors for the same posts. Similarly, recent work shows that AI systems can extract fact-checking information from user comments to produce contextual notes that outperform standard Community Notes in helpfulness ratings \citep{zhang2025commenotes}.

However, these approaches inherit key limitations associated with automated systems. In particular, they remain susceptible to automated bias, as users may over-rely on AI-generated content \citep{nguyen2018believe}. This concern is compounded by the risk of hallucinations in LLMs, which can lead to the generation of plausible but inaccurate information.

Taken together, these findings point toward a broader conclusion: no single approach, whether fully automated or purely human, can simultaneously achieve high accuracy, scalability, and timeliness. As a result, an increasingly supported view in the literature is that the most effective content moderation strategies combine human oversight with AI assistance \citep{gillespie2020content}. This hybrid approach leverages the complementary strengths of humans and machines, enabling moderation that is both more nuanced and scalable \citep{tsvetkova2024new}. AI systems can process vast volumes of content at speeds unattainable for human moderators, while humans provide contextual understanding, moral reasoning, and sensitivity to cultural and political nuances that remain challenging for AI to replicate. \citet{schmitt2024role} show that models integrating human judgment with AI can outperform either component alone.

However, hybrid systems are not without risks. Users often overestimate the accuracy of LLMs, placing trust in AI-generated assessments even when they are incorrect—a phenomenon known as automation bias or machine heuristics \citep{cummings2017automation, deverna2024fact}. Moreover, some human–AI frameworks risk reinforcing users’ existing beliefs, potentially deepening echo chambers and contributing to political polarisation \citep{Jahanbakhsh2023ExploringMedia}. To mitigate these risks, such systems must be carefully designed to avoid unintentionally amplifying bias and division on social media platforms.

\subsection{Theoretical Background}

The capabilities of generative AI offer promising opportunities for collective intelligence between humans and machines \citep{burton2024large, cui2024ai}. However, how to design hybrid systems that fully leverage both machine and human intelligence remains an open question.

Current AI systems are typically designed to be helpful, compliant, and often sycophantic. Empirical evidence shows that AI models affirm users at a much higher rate than humans \citep{cheng2026sychophantic}. This tendency is further amplified in human–AI interactions, where users can steer models toward responses that align with their prior beliefs rather than the truth \citep{sharma2023towards}. In hybrid systems, such behaviour may reduce users’ engagement and critical thinking, as individuals increasingly rely on AI-generated outputs. This phenomenon, referred to as metacognitive laziness, has been observed across a range of tasks \citep{fan2025beware, kosmyna2025your}.

One way to counteract this reduced cognitive engagement is to actively stimulate critical reasoning. Motivated reasoning theory provides a useful framework for understanding how to counteract this effect. According to Kunda’s framework \citep{kunda1990case}, individuals are driven by both the desire to reach accurate conclusions and the desire to defend preferred beliefs. When confronted with information that contradicts their prior views, individuals tend to engage in more effortful and critical processing \citep{ditto1992motivated}.

Building on this insight, introducing controlled bias into AI outputs may enhance human engagement in hybrid systems. Rather than reinforcing users’ existing beliefs, deliberately biased AI can present non-aligned or opposing perspectives, prompting users to scrutinise information more carefully and defend their reasoning. Supporting this idea, recent work shows that collaboration with a partisan AI can improve users’ ability to evaluate the veracity of news headlines when exposed to opposing views. However, this comes at a cost, as participants tend to perceive such AI systems as less trustworthy \citep{lai2025based}.

Furthermore, introducing AI systems that present perspectives differing from those of human collaborators can serve as a mechanism for incorporating political diversity into the moderation process. Diversity of information in problem-solving has been shown to improve the quality of outcomes and reduce errors \citep{hong2004groups, yaniv2011group}. Building on this, a substantial body of research suggests that political diversity can be a valuable asset in collaborative environments. One of the key strengths of Community Notes lies in its ability to harness ideological diversity to produce more balanced and neutral notes \citep{kim2025differential}.

Similar dynamics have been observed in other large-scale collaborative systems. For example, research on Wikipedia attributes the platform’s success to the diversity of viewpoints among editors, supported by structured policies and governance mechanisms \citep{arazy2011information, yasseri2012dynamics, shi2019wisdom, yasseri2025computational}. \citet{shi2019wisdom}, in particular, find that higher levels of polarisation among editors are associated with higher-quality articles, especially on politically sensitive topics. Moreover, exposure to cross-partisan perspectives has been shown to reduce both affective and issue-based polarisation \citep{combs2023reducing, argyle2023leveraging}. These findings suggest that, when properly structured, political diversity not only does not hinder collaboration but can also enhance it.

Extending these insights to AI systems, recent work shows that politically diverse LLM agents can similarly converge toward more accurate beliefs through deliberation \citep{chuang2023wisdom}. In human–AI collaborations, designing AI systems to introduce complementary biases has been shown to facilitate the discovery of novel solutions and improve overall task performance \citep{brinkmann2022hybrid}.

In the context of Community Notes, an AI assistant that provides argumentative feedback can serve as a form of “biased AI” that triggers more effortful cognitive processing, prompting contributors to reassess, refine, and strengthen their notes. In addition, such feedback introduces opposing viewpoints, effectively simulating political diversity within the note-writing stage. As a result, argumentative feedback is expected to lead contributors to produce higher-quality notes.

\subsection{Research Objectives}

Overall, existing moderation approaches face a fundamental trade-off: community-based systems such as Community Notes preserve human judgment but are constrained by partisanship and delays, while automated approaches offer scalability but raise concerns about bias, overreliance on machine-generated judgments, and a lack of transparency, which may undermine user trust. This highlights the need for frameworks that effectively combine human oversight with the efficiency of AI systems.

Building on this theoretical background, we propose a human–AI collaboration framework that uses AI-generated feedback to simulate political diversity in content moderation. The goal is to combine the efficiency of AI with the interpretive strengths of human contributors, while ensuring that humans retain final decision-making authority. We investigate whether AI-generated feedback can emulate cross-partisan collaboration and enhance the quality of contributions on the Community Notes platform. Our focus is on how users revise their notes in response to different types of AI-generated feedback. This work contributes to the growing body of research on the dynamics of Human–AI collaboration, especially in sensitive domains such as content moderation. Specifically, this study addresses the following research questions:

\begin{itemize}
\item \textbf{RQ1:} How does the perceived quality of notes vary by the type of feedback received (argumentative, supportive, or neutral)?
\item \textbf{RQ2:} How does the impact of feedback on perceived note quality differ when participants evaluate co-partisan versus cross-partisan posts?
\item \textbf{RQ3:} How does the effect of feedback on perceived quality change based on whether participants believe it was generated by a human expert or an AI model?
\end{itemize}

To investigate these research questions, we conducted an experiment modelled after the Community Notes process. Participants ($N=893$) were shown a post and asked to write a note providing additional context. Each note was then submitted to the GPT-4 language model, which was prompted to generate one of three types of feedback: argumentative, supportive, or neutral. 

Argumentative feedback offers counterpoints to the original note, supportive feedback reinforces the note’s arguments, and neutral feedback simply restates its content. We intentionally did not prompt the LLM to adopt a partisan stance (i.e., Democrat or Republican); instead, the prompt focused solely on the arguments in the note, thereby avoiding stereotyping of either the AI or the participants.

While the language model generated all the feedback, we randomly instructed about half of the participants in each group that a human expert had written the feedback. This manipulation allows us to examine the roles of human favouritism \citep{morewedge2022preference} and automation bias \citep{cummings2017automation} in shaping participants’ responses.

\begin{figure}[H]
  \centering
      \includegraphics[width=\textwidth]{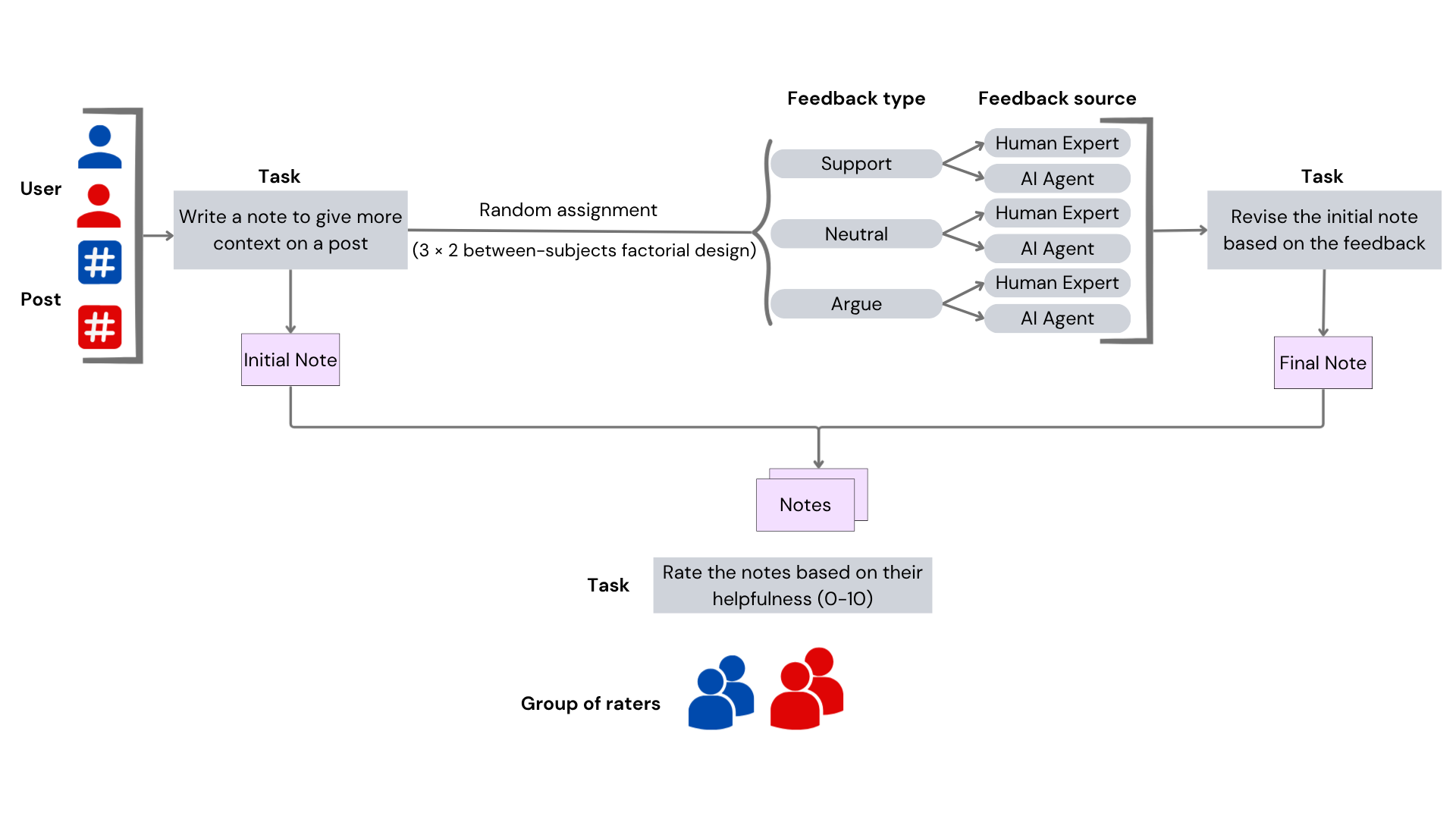}
\caption{\textbf{The experimental workflow illustrates the process of note creation, feedback assignment, and evaluation.}  
Participants (self-identified as Democrats or Republicans) wrote initial notes to provide context on posts authored by either Democrats or Republicans. They then received randomly assigned feedback that varied in type (supportive, neutral, or argumentative) and in source label (AI agent or human expert). After revising their notes in response to the feedback, a separate group of self-identified Democrats and Republicans rated the helpfulness of the original and revised notes. This design enables analysis of feedback effects on note quality and partisan differences in evaluation.}
\label{fig:experiment_diagram}
\end{figure}

Participants were then asked to revise their notes in light of the feedback they received. In a second experiment ($N=1354$), we crowdsourced evaluations of note quality before and after revision. Self-identified Republicans and Democrats rated the notes on a 0–10 scale for helpfulness. Figure \ref{fig:experiment_diagram} provides a schematic overview of the experimental workflow.

The argumentative AI is designed to function as a form of “biased AI”, providing counterarguments to the participants’ initial notes. This design aimed to encourage participants to leverage their full cognitive resources and to consider opposing viewpoints. Consistent with prior research \citep{lai2025based}, participants were generally less likely to engage with argumentative AI feedback. However, among participants who engaged with the AI feedback, argumentative feedback yielded the highest likelihood of improvement. 

Overall, our results highlight the potential of AI to assist humans in sensitive tasks such as content moderation. They further illuminate the mechanisms of collective hybrid intelligence, demonstrating that the productivity of Human-AI teams depends critically on both the extent to which humans actively engage with the AI and the type of feedback it provides. Additionally, our findings suggest that Human-AI collaboration can effectively simulate political diversity in the note-writing process, thereby enhancing the overall performance of Community Notes.

\section{Results}

\subsection{Descriptive Results}
\begin{figure}[H]
    \centering

       \hspace{-0.05 \textwidth} \includegraphics[width=1.1\textwidth]{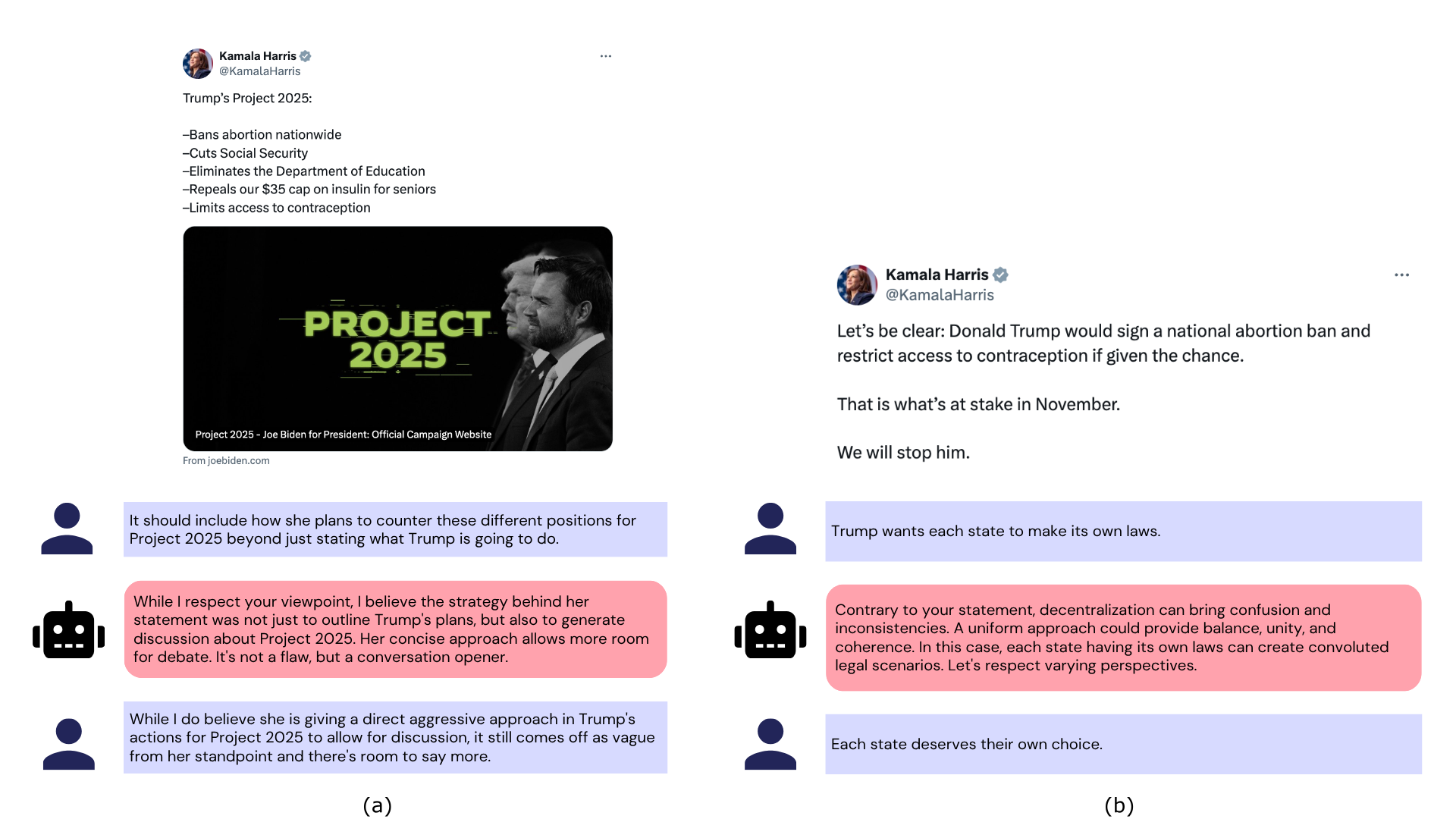}
    \caption{\textbf{Examples of notes with high and low engagement in response to argumentative feedback.}  
(a) A note that improved following argumentative feedback.  
(b) A note that declined following argumentative feedback}
    \label{fig:example_improved}
\end{figure}

Our results show that participants engaged with AI-generated feedback to varying degrees. Figure \ref{fig:example_improved} presents two contrasting cases in which participants received argumentative feedback from the AI agent. In Figure \ref{fig:example_improved}(a), the participant maintained their original stance but incorporated phrasing from the feedback into the revised note. In contrast, in Figure \ref{fig:example_improved}(b), the participant disregarded the feedback entirely, making no meaningful changes.

Comparing how these revised notes were evaluated by Democrat and Republican raters reveals a clear pattern. The note in Figure \ref{fig:example_improved}(a), which engaged with the feedback, received higher ratings. Conversely, the note in Figure \ref{fig:example_improved}(b), which ignored the feedback, was rated poorly.

These contrasting outcomes highlight a key insight: the extent to which participants engage with AI-generated feedback significantly influences the effectiveness of their revisions.

\subsection{Empirical Findings and Statistical Results}

\begin{figure}[H]
    \centering

        \includegraphics[width=.6\textwidth]{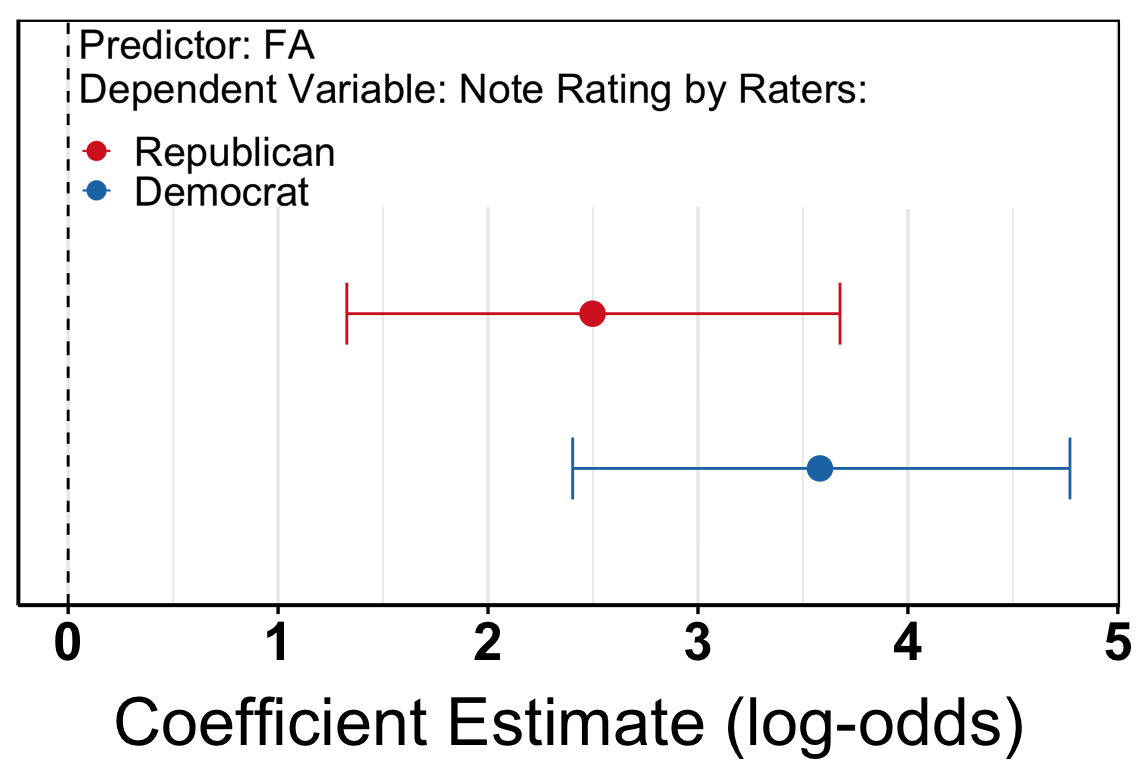}
    \caption{\textbf{Logistic regression for the note rating improvement.}  
Coefficient estimates from the logistic regression model for note rating improvement as a function of $FA$. Points represent estimated log-odds, with horizontal bars indicating 95\% confidence intervals. A vertical dashed line at 0 denotes the null effect. Estimates are grouped by rater affiliation, with blue indicating Democrats and red indicating Republicans. }
    \label{fig:logit_for_I_with_FA}
\end{figure}

To measure engagement, we introduce the Feedback Acceptance rate ($FA$), a metric that quantifies the extent to which participants incorporated the provided feedback into their final submissions. Specifically, $FA$ is defined as the semantic similarity score between the feedback and the revised note (see Methods for details). To evaluate whether engagement with feedback leads to higher-quality revisions, we use $FA$ as a predictor in ordinal logistic regression models. The dependent variables indicate whether the revised note declined, remained unchanged, or improved according to the Democrat raters ($I_\mathrm{D}$) and the Republican raters ($I_\mathrm{R}$). The results (Figure \ref{fig:logit_for_I_with_FA}) show that $FA$ is a strong and consistent predictor of note improvement as perceived by both partisan groups. A comparison of coefficients further indicates that $FA$ has a slightly stronger effect on improvement ratings from Democrats than from Republicans (log-odds for $I_{\mathrm{D}} = 3.581$, $p<0.001$; log-odds for $I_{\mathrm{R}} = 2.498$, $p<0.001$). The regression results are available in Table S1.

\begin{figure}[H]
    \centering

        \includegraphics[width=\textwidth]{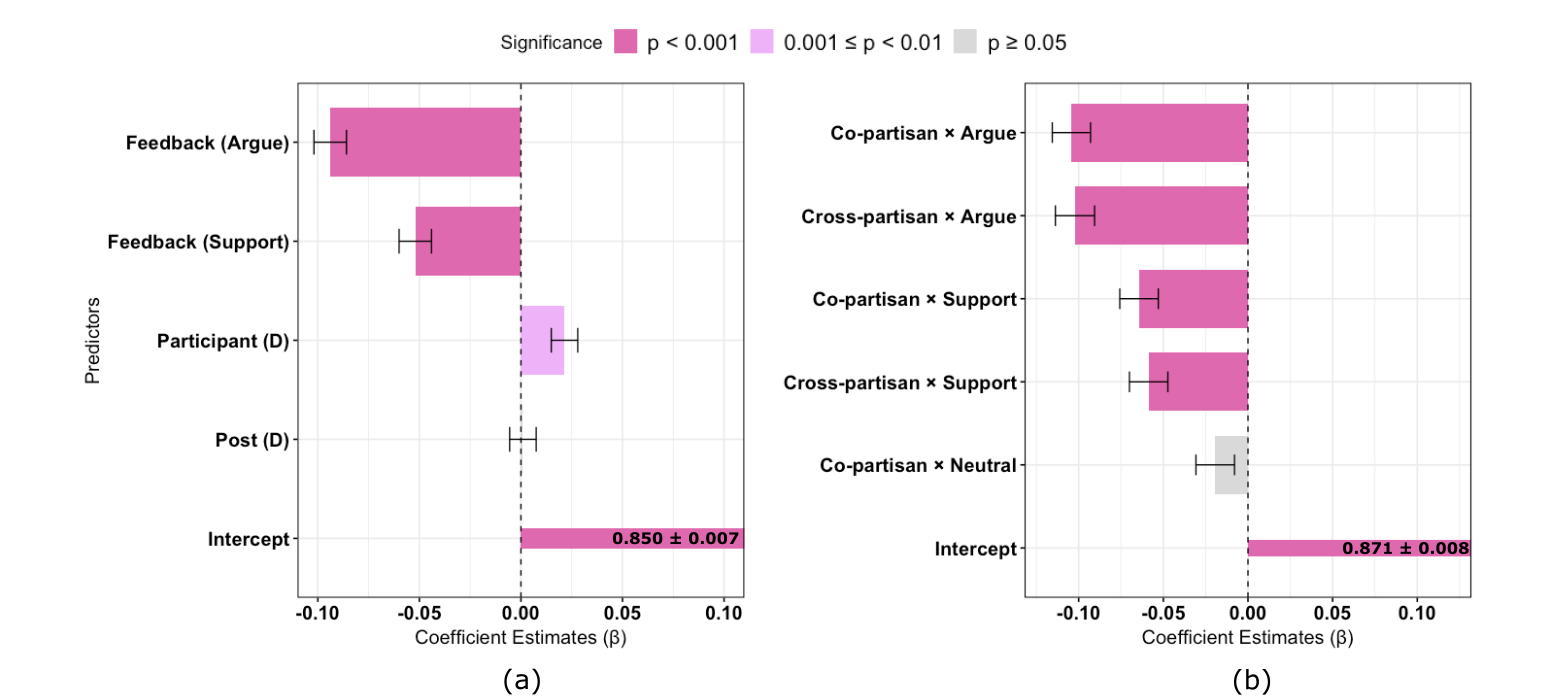}
      \caption{\textbf{Bar plot of OLS model for $FA$.} 
(a) Model 1 includes feedback type (Neutral as the reference), post type (Republican as the reference), and participant partisanship (Republican as the reference).
(b) Model 2 includes feedback type, the alignment between participant and post-partisanship (Co- vs Cross-partisan), and their interaction, with Cross-partisan × Neutral as the reference category. The plots show estimated regression coefficients ($\beta$) with one-standard-error bars. The vertical dashed line at 0 indicates the absence of an effect. Bar widths reflect the relative magnitude of coefficients, and fill colours indicate statistical significance (two-sided Wald tests). }
   
    \label{fig:ols_FA}
\end{figure}

To better understand the factors that influence this engagement metric, we analyse the relationship between $FA$ and three key variables: the type of post (supporting either Democratic or Republican ideology), the participant’s self-identified partisanship, and the type of feedback received. The results from the Ordinary Least Squares (OLS) regression are shown in Figure \ref{fig:ols_FA} (a).

The analysis indicates that users who received argumentative feedback were less likely to incorporate it into their final notes compared to those who received supportive feedback (Argue: $\beta = -0.094$, $p < 0.001$; Support: $\beta = -0.052$, $p < 0.001$; neutral feedback as the reference category). Additionally, Democratic participants were more likely than Republicans to incorporate feedback into their final notes ($\beta = 0.021$, $p = 0.001$; Republican partisanship as the reference category). In contrast, the type of post did not have a significant effect on this behaviour ($p = 0.880$). See Table~S2 for more details.

However, analysing the type of post in isolation may not provide the complete picture. We expect individuals who share the political ideology of the post's author to respond differently when moderating content compared to those with opposing views. To investigate this, we introduce a binary alignment variable that captures the relationship between a participant’s self-identified partisanship and the post’s stance: co-partisan if the participant and the post share the same ideology; cross-partisan otherwise. A post is classified as co-partisan when it supports Democrats (or Republicans) and the participant identifies as a Democrat (or Republican); otherwise, it is classified as cross-partisan.

To examine how political alignment affects engagement with feedback, we conducted an OLS regression using $FA$ as the dependent variable, including an interaction term between co-partisanship and feedback type. The results, shown in Figure \ref{fig:ols_FA} (b), reveal no significant differences in how users incorporate feedback into their notes, regardless of whether they share the same political affiliation as the post. This holds for both supportive and argumentative feedback (See Table S3 for more details).

\begin{figure}[H]
  \centering
      \includegraphics[width=\textwidth]{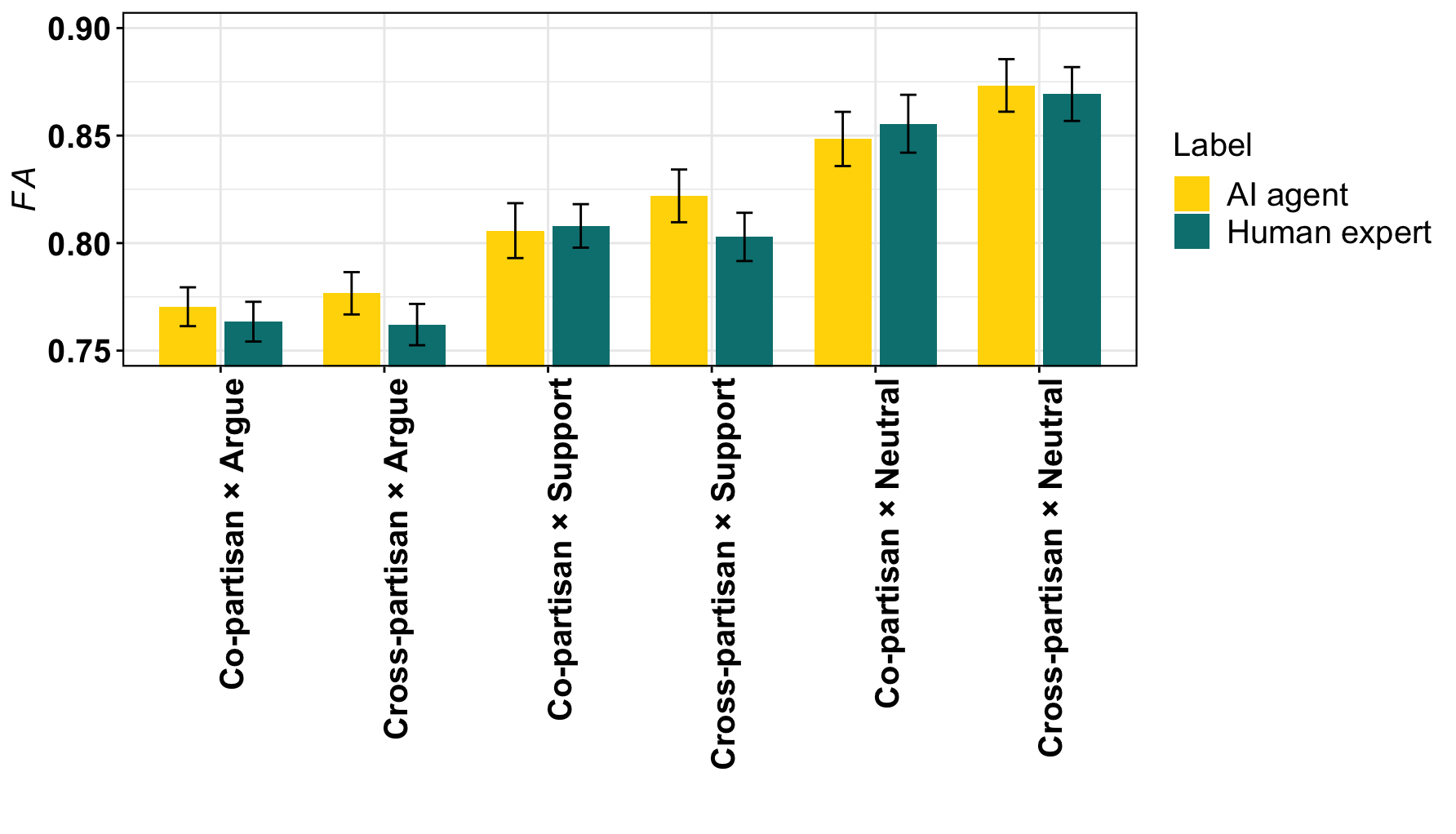}
\caption{\textbf{Comparison of $FA$ values by treatment condition and feedback source for different source labels.} Mean $FA$ values are shown for each feedback type and the alignment between the participant and the post partisanship (Co- vs\ Cross-partisan), with bars coloured by the feedback source (AI vs\ Human). Error bars represent the standard error of the mean.}

\label{fig:fbu_ols_AI and human}
\end{figure}

The second treatment condition manipulated the labelled source of the feedback. Specifically, we investigate whether participants respond differently depending on whether they believe the feedback was generated by a human expert or an AI agent. To explore this, we replicate the previous regression analysis on two separate subsets of the data: one in which feedback was attributed to an AI agent, and another in which it was attributed to a human expert.
Figure \ref{fig:fbu_ols_AI and human} presents the values of $FA$ across the different treatment conditions for both feedback source labels. While the overall patterns are consistent across conditions, we observe a slight increase in $FA$ when the feedback is labelled as AI-generated, specifically for cross-partisan posts. This pattern holds for both argumentative and supportive feedback. These results suggest a marginal preference for AI-labelled feedback when participants engage with content that opposes their political views. However, the differences are not statistically significant (e.g.,$FA$ for AI in cross-partisan × argumentative = 0.777 vs 0.762 for human expert, $p = 0.29$; $FA$ for AI in cross-partisan × supportive = 0.822 vs. 0.803 for human expert, $p = 0.253$), indicating that the labelled source of feedback does not meaningfully influence user behaviour in this context. See Tables~S7 and S8 for more details.

To construct the complete model and evaluate the effect of feedback type, conditional on participant engagement, we estimated an ordinal logistic regression that included both the main effects of feedback type and its interaction with $FA$. The dependent variables are $I_\mathrm{D}$ and $I_\mathrm{R}$, which reflect the improvement of the notes according to Democrat and Republican raters, respectively. This model enables us to determine whether argumentative or supportive feedback is more likely to lead to improvements in perceived helpfulness when participants actively engage with the feedback, and whether these effects vary across different political subgroups.

\begin{figure}[H]
    \centering

        \includegraphics[width=\textwidth]{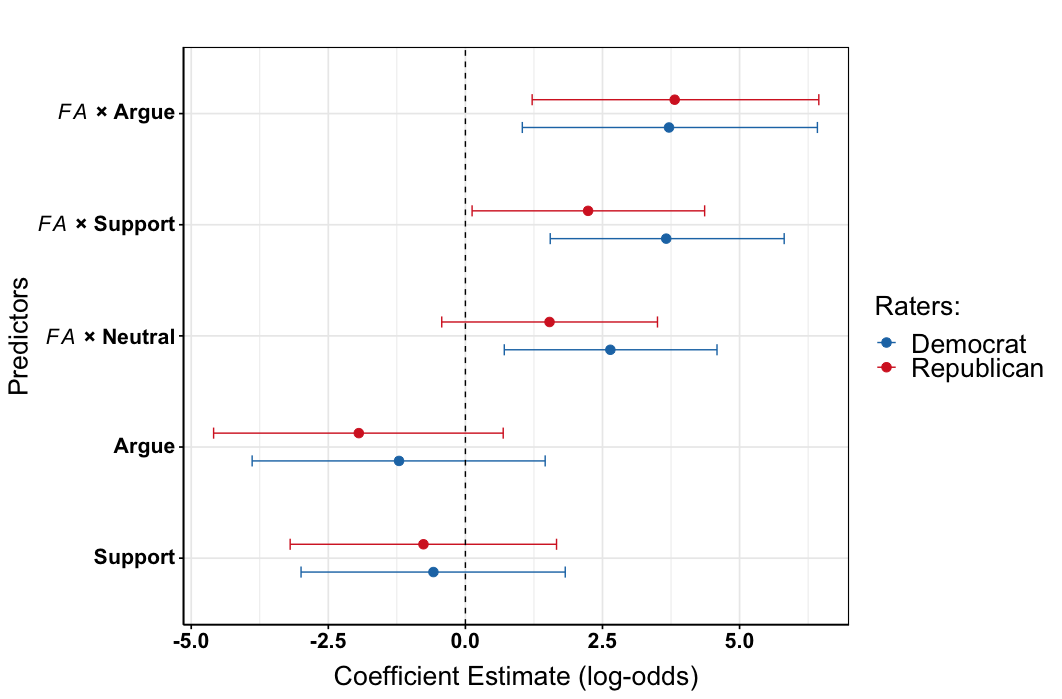}
    \caption{\textbf{Logistic regression for the note rating improvements based on feedback type and $FA$.}  
Coefficient estimates from a logistic regression model predicting $I_\mathrm{D}$ and $I_\mathrm{R}$ using Feedback Acceptance rate ($FA$), feedback type (with \textit{Neutral} as the reference), and their interaction. Points represent estimated coefficients (log-odds), with horizontal bars indicating 95\% confidence intervals. A vertical dashed line at 0 denotes the null effect. Estimates are grouped by rater affiliation, with blue indicating Democrats and red indicating Republicans. }
    \label{fig:logit_I_FA_interaction_with_feedback}
\end{figure}

The results of this analysis are presented in Figure~\ref{fig:logit_I_FA_interaction_with_feedback}. The findings indicate that once participants incorporate feedback into their notes, both argumentative and supportive feedback increase the likelihood of producing a more helpful note, as rated by both Democratic and Republican evaluators. However, argumentative feedback shows a larger effect size than supportive feedback, as assessed by both Democrat raters (Feedback (Argue) $\times$ $FA$: log-odds $ = 3.713$, $p = 0.006$; Feedback (Support) $\times$ $FA$: log-odds $= 3.662$, $p < 0.001$) and Republican raters (Feedback (Argue) $\times$ $FA$: log-odds $ = 3.817$, $p = 0.004$; Feedback (Support) $\times$ $FA$: log-odds = $2.237$, $p = 0.038$). See Table~S4 for more details.

To ensure robustness, we estimate a range of alternative model specifications, with detailed results reported in Tables~S5, S6, S9, S10, S11, S12, S13. We also test the proportional odds assumption underlying the ordinal logistic models using the Brant test. While the assumption holds for Republican ratings, it is rejected for Democratic ratings, indicating that some effects may vary across outcome thresholds. To account for this, we estimate multinomial logit models that relax this assumption. The results remain consistent with our main findings (see SI for details).

\section{Discussion}
As Community Notes gains traction across major social media platforms as a leading approach to crowd-based content moderation \citep{meta2025, presser2025tiktok}, concerns have been raised about its efficiency and long-term sustainability \citep{allen2022birds, chuai2024did, truong2025delayed, augenstein2025community}. One emerging response to these challenges is the integration of generative AI, particularly large language models (LLMs), into the Community Notes ecosystem. 

In this study, we introduce a human–AI collaboration framework in which GPT-4 provides structured feedback on user-generated notes. This approach allows us to both evaluate the effectiveness of AI-generated feedback and examine the mechanisms through which such feedback influences user behaviour. Our findings show that improvements in note quality are not driven solely by the type of feedback received, but critically depend on the extent to which participants engage with that feedback.

\begin{figure}[H]
    \centering

        \includegraphics[width=0.65\textwidth]{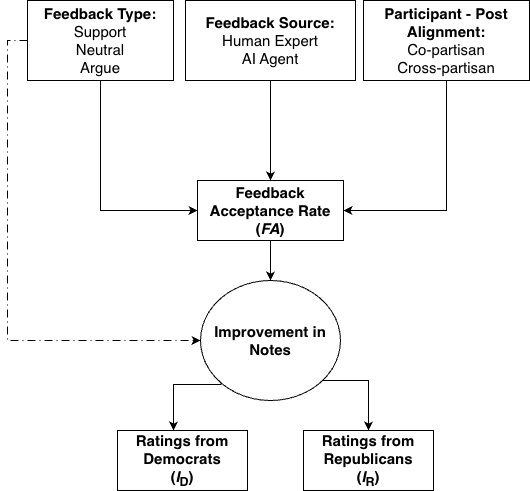}
    \caption{\textbf{Conceptual framework of the study.}  
Feedback type, feedback source label, and participant–post partisan alignment influence participants’ engagement with the feedback, measured by the Feedback Acceptance rate ($FA$). In turn, $FA$ affects improvements in note helpfulness, evaluated separately by Democratic and Republican raters. Overall, feedback type has an indirect effect on note improvement through its interaction with $FA$.}
    \label{fig:conceptual_framework}
\end{figure}

Studies have shown that two essential factors in collective intelligence are the effort a group invests and the effectiveness with which members leverage each other’s knowledge and skills \citep{woolley2024understanding}. Our findings suggest that human-AI collective efforts follow the same principle: participants who engaged with AI-generated feedback more and incorporated it into their revisions produced notes that were perceived as better. Specifically, the Feedback Acceptance rate ($FA$), which reflects how closely participants’ final notes match the feedback, emerged as a significant predictor of note improvement according to both Democratic and Republican raters ($I_\mathrm{D}$ and $I_\mathrm{R}$). A high $FA$ indicates strong engagement with the feedback, while a low $FA$ often corresponds to users disregarding the feedback entirely, sometimes resulting in notes of lower quality than the original (See Figure S10). $FA$ is lower once the feedback is argumentative; This pattern mirrors behaviours observed on social media, where users often respond defensively to content that challenges their political beliefs \citep{bail2018exposure}, and aligns with prior research showing that users distrust AI systems offering opposing viewpoints \citep{lai2025based}. Nevertheless, our findings suggest that when participants move beyond this reflexive partisan response and actively engage with feedback, they produce higher-quality notes. Figure~\ref{fig:conceptual_framework} summarises the conceptual framework underlying our analysis. 

Among participants who actively engaged with the feedback, our findings indicate that argumentative feedback was more likely to result in improvements in note quality. This may be because engaging with counterarguments encourages users to reconsider and refine their positions, resulting in notes that are more balanced, better justified, and ultimately more helpful. Prior research suggests that in problem-solving contexts, when AI systems are designed to complement human cognitive biases, they can prompt users to discover novel solutions, thereby improving outcomes \citep{brinkmann2022hybrid}. Similarly, a biased AI assistant can motivate humans to challenge competing views and increase their cognitive effort \citep{lai2025based}. In our study, the argumentative AI functioned as such a biased assistant, providing an opposing perspective that nudged participants to consider alternative viewpoints and thereby introduced diversity into the note-writing process. Relatedly, \citet{juncosa2026benefit} demonstrate that diversity among groups of note authors can enhance note quality under certain conditions. Our current findings suggest that argumentative AI feedback may serve as a form of political diversity, exposing users to alternative viewpoints and, in turn, fostering higher-quality content.

In our second research question, we investigated how the political alignment between participants and posts affects the perceived helpfulness of notes. Our findings indicate no difference in how users incorporate feedback when the post aligns with their political ideology (co-partisan) versus when it does not (cross-partisan). This pattern holds for both argumentative and supportive feedback. Because the Feedback Acceptance rate ($FA$) is unaffected by participant-post alignment, note improvement is similarly unaffected.

Lastly, in our third research question, we examined how the labelled source of feedback, whether an AI agent or a human expert, influences participant behaviour and, consequently, the quality of their notes. We find a slight preference for AI-labelled feedback when the post comes from the opposing political party (i.e., in cross-partisan contexts), though this difference was not statistically significant. In these cases, participants are more likely to engage with feedback they believe was generated by an AI agent. This finding is consistent with prior research on automated content moderation, which suggests that users tend to be more receptive to algorithmic decisions. This tendency, known as automation bias, reflects the inclination to overestimate the neutrality and competence of automated systems, even when they make errors \citep{cummings2017automation, nguyen2018believe}.

This study offers insights into the evolving role of AI in crowd-based content moderation. We demonstrate that integrating a large language model (LLM) into the Community Notes system can enhance its efficiency, provided the design preserves human agency and the platform’s core participatory nature. In our framework, the AI functions solely as a reviewer offering suggestions. At the same time, users retain complete control over the content of their notes, helping to prevent the imposition of any algorithmic bias.

A key contribution of our approach is introducing political diversity at the note-writing stage, rather than relying solely on diverse raters at the evaluation stage. When participants engage with argumentative feedback, they are more likely to produce higher-quality notes, as rated by both Democratic and Republican evaluators. This suggests that such feedback not only promotes quicker production of helpful notes but also encourages users to consider counterarguments and revise their contributions in a more balanced and thoughtful way, ultimately discouraging the reinforcement of partisan narratives. Compared to pairing two politically diverse human participants to co-author a note, this approach is also less resource-intensive. However, realising the benefits of such hybrid collective intelligence requires human participants to actively engage with the AI, much as collective intelligence emerges in human groups.

Our findings shed light on the underlying mechanisms through which hybrid collective intelligence emerges in human–AI collaboration. On the one hand, engagement with AI feedback is crucial. Without active engagement, the benefits of AI feedback cannot be realised, just as collective intelligence among humans only works when they collaborate. On the other hand, the argumentative nature of the feedback plays a central role in fostering political diversity and balance at the note-writing stage. By prompting participants to confront counterarguments and revise their notes accordingly, argumentative feedback encourages deeper deliberation and reduces the risk of partisan echo chambers. Together, these mechanisms illustrate how hybrid collectives can leverage AI not as a replacement for human reasoning, but as a catalyst that enhances the quality and inclusiveness of collective outputs.

While our findings highlight the potential of integrating argumentative AI feedback into Community Notes, several important limitations must be acknowledged. First, this study does not assess the objective factual accuracy of the notes. Instead, note quality is evaluated through crowdsourced ratings, with groups of Democrats and Republicans assessing perceived helpfulness. This approach does not provide a direct measure of truthfulness; for example, professional fact-checkers might arrive at different conclusions. However, our method aligns with the existing structure of Community Notes, where note approval and display are determined by community consensus rather than expert adjudication. \citet{augenstein2025community} describe this as an epistemological distinction between traditional fact-checking and Community Notes, where traditional fact-checking seeks objective truth verified by experts, whereas Community Notes reflect subjective judgments based on users’ perspectives. Importantly, the aim of Community Notes is not solely to debunk falsehoods but to add context to potentially misleading posts. These posts may not contain outright inaccuracies, yet still risk misinforming readers through omission or imbalance. In such cases, a politically diverse and balanced group of laypeople is well-suited to assess whether the contextual information provided is genuinely informative and helpful \citep{martel2024crowds}. Studies have found substantial agreement with the Community Notes and professional fact-checkers \citep{saeed2022crowdsourced}. 

Second, Content moderation encompasses a broad range of practices, including detecting hate speech, harassment, spam, and coordinated inauthentic behaviour. Within this broader landscape, identifying and contextualising potentially misleading political content represents only one component. However, the consequences of such content can be particularly severe, as the spread of political misinformation has been shown to undermine public trust and threaten democratic processes \citep{greene2021misremembering, klein2025mobilising}. In this study, we focus on how the hybrid collective intelligence of humans and AI can improve the identification and contextualisation of such content. While our empirical setting is specific to potentially misleading political content in the U.S. context, the underlying mechanisms we identify, particularly the role of engagement and exposure to diverse perspectives, are likely to generalise to other areas of content moderation where interpretation, judgment, and user interaction are central.

Third, findings derived from an experimental setting may not fully generalise to real-world contexts. For instance, Community Notes participants are volunteers often motivated by political or ideological commitments, whereas participants in our study were recruited and compensated. This difference is reflected in the tone and structure of the notes produced here, which diverge from those typically found on the Community Notes platform. Additionally, in community-based efforts, a small group of ``power users'' often comes to dominate content creation, developing a specialised language and tone tailored to the platform’s norms. This phenomenon is evident in Community Notes, where the widespread use of shorthand, such as “NNN” (“No Note Needed”), signals that a post is not misleading and does not require annotation. New or casual users are unlikely to be familiar with such conventions, resulting in experienced contributors acting as gatekeepers, a dynamic that may hinder platform scalability \citep{halfaker2013rise}. It is essential to emphasise that this study did not aim to replicate Community Notes, but rather to investigate the impact of integrating AI-driven argumentative assistance into the community-based content moderation process.

Fourth, the “breaking news problem” remains a persistent challenge. This occurs when a note addresses an event of which the AI is unaware. For AI to provide meaningful feedback, it must have a sufficient understanding of the topic. When this is missing, the AI may still add value by prompting users to think critically—acting like a nudging algorithm \citep{pennycook2022nudging} that encourages them to seek out reliable sources and investigate further. That said, recent advancements in large language models have enabled web search capabilities, which may enable them to engage more effectively with current events in future applications.

Overall, this study provides a deeper understanding of human–AI interaction within a community-based content moderation system. Our findings suggest that AI can meaningfully enhance this process, not by replacing human judgment, but by offering targeted feedback. In particular, AI-generated argumentative feedback can introduce political diversity and improve the quality of notes by prompting authors to consider opposing viewpoints. However, the presence of AI alone is insufficient to ensure effective collaboration. Its impact depends on active user engagement.

This insight is particularly relevant in light of recent developments in Community Notes, where top contributors can use AI note writers to propose a note \citep{x2024communitynotesapi}. A deeper understanding of human–AI interaction in this context is therefore essential for designing systems that preserve human agency while effectively leveraging AI to enhance collective outcomes.

Future work could extend these insights by examining longer-term interactions between users and AI to understand how engagement patterns evolve over time. Allowing users to interact with the AI across multiple sessions would provide a deeper understanding of sustained behaviour in this context. Additionally, while this study focused on political content, it would be valuable to explore how human–AI collaboration functions in apolitical or less polarising domains to assess whether similar engagement patterns emerge.

\section{Methods}
\subsection{Data Collection}
We collected data through two online experiments conducted on the Prolific platform. The study consisted of two phases: (1) a note-writing experiment in which participants created and revised notes, and (2) a separate evaluation experiment in which a new group of participants assessed note quality. We implemented both experiments using the oTree framework (version 5.10.4), and GPT-4 was used as the AI agent for feedback generation.

Participants in both experiments were required to be residents of the United States. Across both studies, we excluded individuals who did not provide consent, selected a neutral political affiliation, did not complete the experiment, or were timed out.

\subsubsection{Note Writing}
\textbf{Participants.} 
We used 40 social media posts with clear partisan messaging, all of which had been previously analysed by Community Notes users (see Supplementary Information for more details). After data cleaning, the final sample comprised 893 participants, including 455 self-identified Democrats and 438 self-identified Republicans. The study received ethical approval from the University College Dublin Office of Research Ethics (HS-LR-24-230-Mohammadi-Yasseri).

\noindent
\textbf{Procedure.}
Participants first reported their political affiliation using a nine-point scale ranging from ``Strongly Democrat'' to ``Strongly Republican''. Those who identified as neutral were excluded. We then showed participants a political social media post and asked them to write a contextual note and classify the post as either misleading or not. The instructions mirrored those provided to Community Notes contributors. Participants were given 10 minutes to complete this initial note-writing task.

After submitting their note, we showed participants feedback labelled as coming from either an AI agent or a human expert, although all feedback was generated by GPT-4. We then gave the participants 7 minutes to review the feedback and revise their notes. They were also allowed to update the post's classification. Finally, participants completed a short survey measuring their comfort with emerging technologies. At the end of the experiment, we debriefed the participants about the true source of the feedback. See Supplementary Information (SI) for more details about the experiment procedure. 

\begin{figure}
  \centering
      \includegraphics[width=\textwidth]{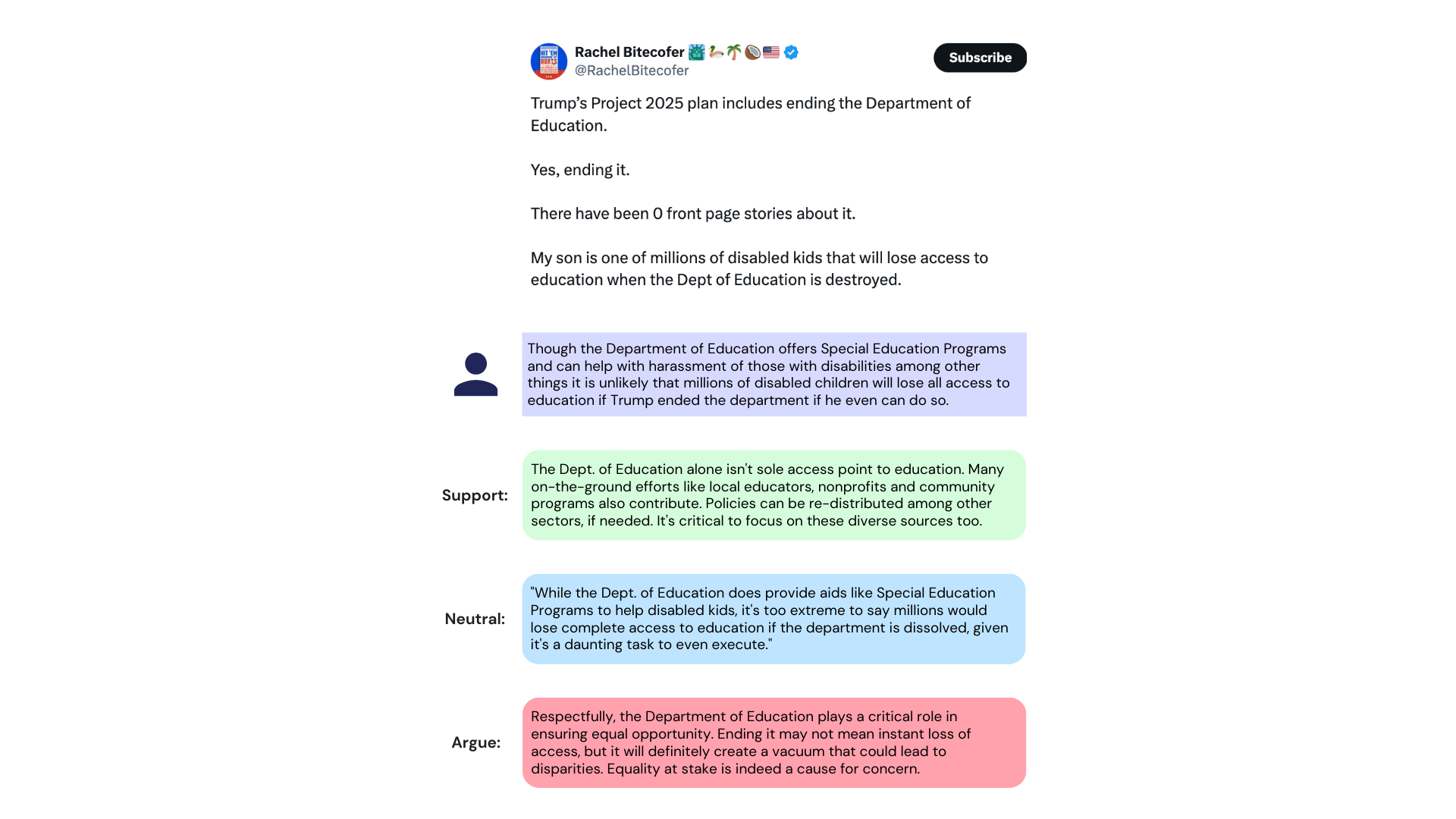}
\caption{\textbf{AI-generated feedback across treatment conditions.} An example of a note created in the experiment and the responses the AI provides in each treatment.}
\label{fig:experiment_treatments}
\end{figure}

\noindent
\textbf{Experimental Conditions.} We randomly assigned participants to a 3 × 2 between-subjects design. The first factor manipulated the type of feedback provided: Argue, Neutral, or Support. In the Argue condition, the AI-generated feedback challenged the participant’s note and presented opposing arguments. In the Neutral condition, the AI rephrased the note without introducing new arguments or contradictions. In the Support condition, the AI reinforced and expanded upon the participant’s original arguments. Examples of feedback across conditions are shown in Figure \ref{fig:experiment_treatments}.

The second factor manipulated the perceived source of the feedback. We randomly informed participants that the feedback was generated either by a human expert or by an AI agent, with approximately half assigned to each condition. In reality, all feedback was generated by GPT-4.

\noindent
\textbf{Feedback Generation.} We generated feedback using the OpenAI API (version 1.25.0) via the \texttt{chat.completions} endpoint, with the model set to \texttt{gpt-4}. The temperature parameter was set to 1.0 to allow for variability in responses, while all other parameters were left at their default values.  

We designed separate prompts for each feedback condition (argue, support, and neutral), along with a shared system message. The full verbatim prompts and system instructions are provided in the ``Feedback Prompts'' section in Supplementary Information (SI).

\noindent
\textbf{Design Constraints.}  
We implemented a quota-based sampling design to ensure balanced partisan representation. Specifically, each social media post was evaluated, within each treatment condition, by at least three self-identified Democrats and three self-identified Republicans.

\subsubsection{Note Evaluation}
\textbf{Participants.}  
In the second experiment, we recruited a new sample of 1354 participants to evaluate the notes. This study  received ethical approval from the University College Dublin Office of Research Ethics (HS-LR-24-188-Mohammadi-Yasseri).

\noindent
\textbf{Procedure.}  
After providing informed consent, participants reported their political affiliation using a nine-point scale. Those selecting a neutral affiliation were excluded. We then showed the participants two social media posts, each accompanied by 10 notes, and asked them to evaluate the notes. Each participant evaluated a total of 20 notes (10 per post). Participants were given up to 7 minutes per post to complete their evaluations. See Supplementary Information (SI) for more details about the experiment procedure. 

\noindent
\textbf{Measures.}  
We asked participants to evaluate each note along two dimensions: helpfulness and favouritism. Helpfulness was rated on a scale from 0 to 10, reflecting the perceived quality of the note. Favouritism was measured using a nine-point scale ranging from ``Favouring Democrats'' to ``Favouring Republicans''.

\noindent
\textbf{Design Constraints.}  
Each note was evaluated by at least three self-identified Democrats and three self-identified Republicans to ensure balanced partisan representation (Figure~S11). In addition to the general exclusion criteria, we excluded participants who provided identical ratings across all evaluations, as this indicated a lack of engagement with the task.

\subsection{Measurements}

\subsubsection{Helpfulness}

The data collected from the note evaluation experiment provides the helpfulness scores assigned to each note by participants identified as Democrats and Republicans. As described earlier, each evaluator rates two posts and 10 notes per post. To standardise the scores and account for each user's rating habit, we calculated z-scores for the notes associated with each post rated by a rater. The normalised score \(\hat{H}^T_{u,i}\), representing helpfulness given by user \(u\) to note \(i\) for post \(T\), is calculated as

\[\hat{H}^T_{u,i} = \frac{H^T_{u,i}- \mu^T_u}{\sigma^T_u},\]
\noindent
where \(\mu^T_u\) is the mean of the scores assigned by user \(u\) to post \(T\), and \(\sigma^T_u\) is the standard deviation of these scores.

To ensure reliability, we excluded the highest and lowest ratings for each note, provided the note had been rated by at least five Democrats and five Republicans (evaluated separately). We then calculate the final helpfulness score for each note by averaging the remaining normalised ratings.

Improvement is calculated as the relative percentage change in helpfulness scores from the initial to the final note. The improvement metric \(I^H_X\), where \(X\) indicates Democrat or Republican raters, is defined as:

\[
I^H_X = \frac{H_{\mathrm{final}} - H_{\mathrm{initial}}}{|H_{\mathrm{initial}}| + \epsilon},
\]

\noindent where $\epsilon$ is a small constant added to avoid division by zero.

To categorise notes, we first assign the direction of change using the sign of $I^H_X$. To avoid classifying very small changes as meaningful improvement or decline, we apply a magnitude threshold based on the log-transformed absolute change. Specifically, notes are classified as \textit{unchanged} if

\[
\log_{10}(|I^H_X|) < -0.1.
\]

\noindent Otherwise, notes are classified as \textit{improved} ($I = +1$) if $I^H_X > 0$, and \textit{declined} ($I = -1$) if $I^H_X < 0$.

\subsubsection{Feedback Acceptance rate ($FA$)}

To quantify the extent to which participants incorporated the feedback into their final notes, we calculated the semantic similarity between each piece of feedback and the corresponding revised note. This was measured using cosine similarity, implemented in Python with the Scikit-learn package (version 0.0.post9). The resulting metric, referred to as the Feedback Acceptance rate ($FA$), captures the degree to which the feedback has been incorporated in the revised note. Figure S8 displays the distribution of $FA$ scores across the different feedback conditions.

The same semantic similarity method was also used to compute the similarity between the initial and revised versions of each note. This measure captures the extent to which a note changed during the revision process. Lower similarity scores indicate greater changes. Figure S9 shows the distribution of similarity scores between the initial and final notes across the different feedback conditions. 

\subsection{Statistical Analysis}
We conducted ordinal logistic regressions to examine changes in helpfulness scores, which could take values of 1 (``improved''), 0 (``unchanged''), or -1 (``declined''). The regressions were performed using the polr() function from the MASS package (version 7.3.57) in R. We report log-odds, standard errors, and two-tailed p-values for each predictor. The p-values were calculated using the Wald test. To investigate the variables affecting the Feedback Acceptance rate ($FA$), we used ordinary least squares regressions (OLS). We used the lm() function from the stats package (version 4.2.1) in R to estimate the models. We report coefficient estimates, standard errors, and two-tailed p-values for each predictor. The p-values were calculated using the Wald test. More details on the statistical analysis are available in the Supplementary Information (SI).
\section*{Data Availability}
Data used in the study are available on Harvard Dataverse (doi:10.7910/DVN/SWTACT).

\section*{Declaration of generative AI in scientific writing}

During the preparation of this work, the authors used ChatGPT 4.0 to improve the writing style of this article. After using this tool, the authors reviewed and edited the content as needed and take full responsibility for the content of the publication.

\section*{Acknowledgments}

The authors wish to thank Gabriela Juncosa and 
Nico Mutzner for their valuable comments. This publication has emanated from research supported in part by a grant from Taighde Éireann – Research Ireland under Grant numbers 18/CRT/6049 and IRCLA/2022/3217. TY acknowledges support from Workday, Inc. For the purpose of Open Access, the author has applied a CC BY public copyright licence to any Author Accepted Manuscript version arising from this submission. 

\section*{Author Contribution}

SM and TY designed the experiments and analysed the data. SM conducted the experiments. SM and TY drafted the paper. TY secured the funding and supervised the project. All authors contributed to writing the manuscript and gave ﬁnal approval for publication.

\section*{Competing interests}
The authors declare no competing interests.
\section*{Correspondence}
Correspondence and requests for materials should be addressed to Taha Yasseri: taha.yasseri@tcd.ie.
%Bibliography
% \bibliographystyle{unsrt}  
% \bibliography{references}  

\begin{thebibliography}{}

\bibitem[Agunlejika, 2025]{agunlejika2025ai}
Agunlejika, T. (2025).
\newblock Ai-driven fact-checking in journalism: Enhancing information veracity and combating misinformation: A systematic review.
\newblock {\em Available at SSRN 5122225}.

\bibitem[Allcott and Gentzkow, 2017]{allcott2017social}
Allcott, H. and Gentzkow, M. (2017).
\newblock Social media and fake news in the 2016 election.
\newblock {\em {Journal of Economic Perspectives}}, 31(2):211--236.

\bibitem[Allen et~al., 2022]{allen2022birds}
Allen, J., Martel, C., and Rand, D.~G. (2022).
\newblock Birds of a feather don’t fact-check each other: Partisanship and the evaluation of news in {Twitter}’s {Birdwatch} crowdsourced fact-checking program.
\newblock In {\em {Proceedings of the 2022 CHI conference on human factors in computing systems}}, pages 1--19.

\bibitem[Arazy et~al., 2011]{arazy2011information}
Arazy, O., Nov, O., Patterson, R., and Yeo, L. (2011).
\newblock Information quality in {Wikipedia}: The effects of group composition and task conflict.
\newblock {\em {Journal of Management Information Systems}}, 27(4):71--98.

\bibitem[Argyle et~al., 2023]{argyle2023leveraging}
Argyle, L.~P., Bail, C.~A., Busby, E.~C., Gubler, J.~R., Howe, T., Rytting, C., Sorensen, T., and Wingate, D. (2023).
\newblock Leveraging {AI} for democratic discourse: Chat interventions can improve online political conversations at scale.
\newblock {\em {Proceedings of the National Academy of Sciences}}, 120(41):e2311627120.

\bibitem[Augenstein et~al., 2025]{augenstein2025community}
Augenstein, I., Bakker, M., Chakraborty, T., Corney, D., Ferrara, E., Gurevych, I., Hale, S., Hovy, E., Ji, H., Larraz, I., et~al. (2025).
\newblock Community moderation and the new epistemology of fact checking on social media.
\newblock {\em arXiv preprint arXiv:2505.20067}.

\bibitem[Augenstein et~al., 2019]{augenstein2019multifc}
Augenstein, I., Lioma, C., Wang, D., Lima, L.~C., Hansen, C., Hansen, C., and Simonsen, J.~G. (2019).
\newblock {MultiFC}: A real-world multi-domain dataset for evidence-based fact checking of claims.
\newblock In {\em Proceedings of the 2019 conference on empirical methods in natural language processing and the 9th international joint conference on natural language processing ({EMNLP-IJCNLP})}, pages 4685--4697.

\bibitem[Bail et~al., 2018]{bail2018exposure}
Bail, C.~A., Argyle, L.~P., Brown, T.~W., Bumpus, J.~P., Chen, H., Hunzaker, M.~F., Lee, J., Mann, M., Merhout, F., and Volfovsky, A. (2018).
\newblock Exposure to opposing views on social media can increase political polarization.
\newblock {\em {Proceedings of the National Academy of Sciences}}, 115(37):9216--9221.

\bibitem[Bakshy et~al., 2015]{bakshy2015exposure}
Bakshy, E., Messing, S., and Adamic, L.~A. (2015).
\newblock Exposure to ideologically diverse news and opinion on {Facebook}.
\newblock {\em Science}, 348(6239):1130--1132.

\bibitem[Binns et~al., 2017]{binns2017like}
Binns, R., Veale, M., Van~Kleek, M., and Shadbolt, N. (2017).
\newblock Like trainer, like bot? inheritance of bias in algorithmic content moderation.
\newblock In {\em {Social Informatics: 9th International Conference, SocInfo 2017, Oxford, UK, September 13-15, 2017, Proceedings, Part II 9}}, pages 405--415. Springer.

\bibitem[Brinkmann et~al., 2022]{brinkmann2022hybrid}
Brinkmann, L., Gezerli, D., Kleist, K., M{\"u}ller, T.~F., Rahwan, I., and Pescetelli, N. (2022).
\newblock Hybrid social learning in human-algorithm cultural transmission.
\newblock {\em {Philosophical Transactions of the Royal Society A}}, 380(2227):20200426.

\bibitem[Burton et~al., 2024]{burton2024large}
Burton, J.~W., Lopez-Lopez, E., Hechtlinger, S., Rahwan, Z., Aeschbach, S., Bakker, M.~A., Becker, J.~A., Berditchevskaia, A., Berger, J., Brinkmann, L., et~al. (2024).
\newblock How large language models can reshape collective intelligence.
\newblock {\em {Nature Human Behaviour}}, 8(9):1643--1655.

\bibitem[{{Center for Countering Digital Hate}}, 2024]{CCDH2024}
{{Center for Countering Digital Hate}} (2024).
\newblock Rated not helpful; how {X}’s {Community Notes} system falls short on misleading election claims.

\bibitem[Cheng et~al., 2026]{cheng2026sychophantic}
Cheng, M., Lee, C., Khadpe, P., Yu, S., Han, D., and Jurafsky, D. (2026).
\newblock Sycophantic {AI} decreases prosocial intentions and promotes dependence.
\newblock {\em Science}, 391(6792):eaec8352.

\bibitem[Chuai et~al., 2024]{chuai2024did}
Chuai, Y., Tian, H., Pr{\"o}llochs, N., and Lenzini, G. (2024).
\newblock Did the roll-out of {Community Notes} reduce engagement with misinformation on {X/Twitter?}
\newblock {\em Proceedings of the {ACM} on human-computer interaction}, 8(CSCW2):1--52.

\bibitem[Chuang et~al., 2023]{chuang2023wisdom}
Chuang, Y.-S., Suresh, S., Harlalka, N., Goyal, A., Hawkins, R., Yang, S., Shah, D., Hu, J., and Rogers, T.~T. (2023).
\newblock The wisdom of partisan crowds: Comparing collective intelligence in humans and {LLM}-based agents.
\newblock {\em arXiv preprint arXiv:2311.09665}.

\bibitem[Combs et~al., 2023]{combs2023reducing}
Combs, A., Tierney, G., Guay, B., Merhout, F., Bail, C.~A., Hillygus, D.~S., and Volfovsky, A. (2023).
\newblock Reducing political polarization in the {United States} with a mobile chat platform.
\newblock {\em {Nature Human Behaviour}}, 7(9):1454--1461.

\bibitem[Cui and Yasseri, 2024]{cui2024ai}
Cui, H. and Yasseri, T. (2024).
\newblock {AI}-enhanced collective intelligence.
\newblock {\em Patterns}, 5(11).

\bibitem[Cummings, 2017]{cummings2017automation}
Cummings, M.~L. (2017).
\newblock Automation bias in intelligent time critical decision support systems.
\newblock In {\em Decision making in aviation}, pages 289--294. Routledge.

\bibitem[De et~al., 2025]{de2025supernotes}
De, S., Bakker, M.~A., Baxter, J., and Saveski, M. (2025).
\newblock {Supernotes}: Driving consensus in crowd-sourced fact-checking.
\newblock In {\em Proceedings of the {ACM} on {Web Conference} 2025}, pages 3751--3761.

\bibitem[Demartini et~al., 2020]{demartini2020human}
Demartini, G., Mizzaro, S., and Spina, D. (2020).
\newblock Human-in-the-loop artificial intelligence for fighting online misinformation: Challenges and opportunities.
\newblock {\em {IEEE Data Eng. Bull.}}, 43(3):65--74.

\bibitem[DeVerna et~al., 2024]{deverna2024fact}
DeVerna, M.~R., Yan, H.~Y., Yang, K.-C., and Menczer, F. (2024).
\newblock Fact-checking information from large language models can decrease headline discernment.
\newblock {\em {Proceedings of the National Academy of Sciences}}, 121(50):e2322823121.

\bibitem[Ditto and Lopez, 1992]{ditto1992motivated}
Ditto, P.~H. and Lopez, D.~F. (1992).
\newblock Motivated skepticism: Use of differential decision criteria for preferred and nonpreferred conclusions.
\newblock {\em {Journal of Personality and Social Psychology}}, 63(4):568.

\bibitem[Fan et~al., 2025]{fan2025beware}
Fan, Y., Tang, L., Le, H., Shen, K., Tan, S., Zhao, Y., Shen, Y., Li, X., and Ga{\v{s}}evi{\'c}, D. (2025).
\newblock Beware of metacognitive laziness: Effects of generative artificial intelligence on learning motivation, processes, and performance.
\newblock {\em {British Journal of Educational Technology}}, 56(2):489--530.

\bibitem[Gillespie, 2020]{gillespie2020content}
Gillespie, T. (2020).
\newblock Content moderation, {AI}, and the question of scale.
\newblock {\em {Big Data \& Society}}, 7(2):2053951720943234.

\bibitem[Greene et~al., 2021]{greene2021misremembering}
Greene, C.~M., Nash, R.~A., and Murphy, G. (2021).
\newblock Misremembering {Brexit}: Partisan bias and individual predictors of false memories for fake news stories among brexit voters.
\newblock {\em Memory}, 29(5):587--604.

\bibitem[Halfaker et~al., 2013]{halfaker2013rise}
Halfaker, A., Geiger, R.~S., Morgan, J.~T., and Riedl, J. (2013).
\newblock The rise and decline of an open collaboration system: How {Wikipedia}’s reaction to popularity is causing its decline.
\newblock {\em {American Behavioral Scientist}}, 57(5):664--688.

\bibitem[Hassan et~al., 2015]{hassan2015quest}
Hassan, N., Adair, B., Hamilton, J.~T., Li, C., Tremayne, M., Yang, J., and Yu, C. (2015).
\newblock The quest to automate fact-checking.
\newblock In {\em {Proceedings of the 2015 Computation + Journalism Symposium}}. {The Computation+ Journalism Symposium New York}.

\bibitem[Hoes et~al., 2023]{hoes2023leveraging}
Hoes, E., Altay, S., and Bermeo, J. (2023).
\newblock Leveraging {ChatGPT} for efficient fact-checking.
\newblock {\em PsyArXiv. April}, 3.

\bibitem[Hong and Page, 2004]{hong2004groups}
Hong, L. and Page, S.~E. (2004).
\newblock Groups of diverse problem solvers can outperform groups of high-ability problem solvers.
\newblock {\em {Proceedings of the National Academy of Sciences}}, 101(46):16385--16389.

\bibitem[Jahanbakhsh et~al., 2023]{Jahanbakhsh2023ExploringMedia}
Jahanbakhsh, F., Katsis, Y., Wang, D., Popa, L., and Muller, M. (2023).
\newblock Exploring the use of personalized {AI} for identifying misinformation on social media.
\newblock In {\em Proceedings of the 2023 {CHI} Conference on Human Factors in Computing Systems}, pages 1--27.

\bibitem[Juncosa et~al., 2026]{juncosa2026benefit}
Juncosa, G., Mohammadi, S., Samahita, M., and Yasseri, T. (2026).
\newblock The benefit of collective intelligence in community-based content moderation is limited by overt political signalling.
\newblock {\em arXiv preprint arXiv:2601.22201}.

\bibitem[Kaliyar et~al., 2021]{kaliyar2021fakebert}
Kaliyar, R.~K., Goswami, A., and Narang, P. (2021).
\newblock {FakeBERT}: Fake news detection in social media with a {BERT}-based deep learning approach.
\newblock {\em {Multimedia Tools and Applications}}, 80(8):11765--11788.

\bibitem[Kaplan, 2025]{meta2025}
Kaplan, J. (2025).
\newblock More speech and fewer mistakes.
\newblock Accessed: (February 25, 2025).

\bibitem[Kim et~al., 2025]{kim2025differential}
Kim, J., Wang, Z., Shi, H., Ling, H.-K., and Evans, J. (2025).
\newblock Differential impact from individual versus collective misinformation tagging on the diversity of {Twitter} ({X}) information engagement and mobility.
\newblock {\em {Nature Communications}}, 16(1):973.

\bibitem[King et~al., 2017]{king2017news}
King, G., Schneer, B., and White, A. (2017).
\newblock How the news media activate public expression and influence national agendas.
\newblock {\em Science}, 358(6364):776--780.

\bibitem[Klein, 2025]{klein2025mobilising}
Klein, O. (2025).
\newblock Mobilising the mob: The multifaceted role of social media in the january 6th {US} capitol attack.
\newblock {\em {Javnost-The Public}}, 32(1):33--50.

\bibitem[Kosmyna et~al., 2025]{kosmyna2025your}
Kosmyna, N., Hauptmann, E., Yuan, Y.~T., Situ, J., Liao, X.-H., Beresnitzky, A.~V., Braunstein, I., and Maes, P. (2025).
\newblock Your brain on {ChatGPT}: Accumulation of cognitive debt when using an {AI} assistant for essay writing task.
\newblock {\em arXiv preprint arXiv:2506.08872}, 4.

\bibitem[Kunda, 1990]{kunda1990case}
Kunda, Z. (1990).
\newblock The case for motivated reasoning.
\newblock {\em {Psychological Bulletin}}, 108(3):480--498.

\bibitem[Kuznetsova et~al., 2025]{kuznetsova2025generative}
Kuznetsova, E., Makhortykh, M., Vziatysheva, V., Stolze, M., Baghumyan, A., and Urman, A. (2025).
\newblock In generative {AI} we trust: can chatbots effectively verify political information?
\newblock {\em {Journal of Computational Social Science}}, 8(1):15.

\bibitem[Lai et~al., 2025]{lai2025based}
Lai, S., Kim, J., Kunievsky, N., Potter, Y., and Evans, J. (2025).
\newblock Biased {AI} improves human decision-making but reduces trust.
\newblock {\em arXiv preprint arXiv:2508.09297}.

\bibitem[Lu et~al., 2022]{lu2022effects}
Lu, Z., Li, P., Wang, W., and Yin, M. (2022).
\newblock The effects of {AI}-based credibility indicators on the detection and spread of misinformation under social influence.
\newblock {\em Proceedings of the {ACM} on {Human-Computer Interaction}}, 6(CSCW2):1--27.

\bibitem[Ma et~al., 2025]{ma2025local}
Ma, J., Hu, L., Li, R., and Fu, W. (2025).
\newblock {LoCal}: Logical and causal fact-checking with {LLM}-based multi-agents.
\newblock In {\em Proceedings of the {ACM} on {Web Conference} 2025}, {WWW '25}, page 1614–1625, New York, NY, USA. {Association for Computing Machinery}.

\bibitem[Martel et~al., 2024]{martel2024crowds}
Martel, C., Allen, J., Pennycook, G., and Rand, D.~G. (2024).
\newblock Crowds can effectively identify misinformation at scale.
\newblock {\em {Perspectives on Psychological Science}}, 19(2):477--488.

\bibitem[Mohammadi et~al., 2025]{mohammadi2025birdwatch}
Mohammadi, S., Chinichian, N., Doyal, H., Skutilova, K., Cui, H., d'Errico, M., Grayson, S., and Yasseri, T. (2025).
\newblock From {Birdwatch} to {Community Notes}, from {Twitter} to {X}: four years of community-based content moderation.
\newblock {\em arXiv arXiv:2510.09585}.

\bibitem[Morewedge, 2022]{morewedge2022preference}
Morewedge, C.~K. (2022).
\newblock Preference for human, not algorithm aversion.
\newblock {\em {Trends in Cognitive Sciences}}, 26(10):824--826.

\bibitem[Nguyen et~al., 2018]{nguyen2018believe}
Nguyen, A.~T., Kharosekar, A., Krishnan, S., Krishnan, S., Tate, E., Wallace, B.~C., and Lease, M. (2018).
\newblock Believe it or not: designing a human-{AI} partnership for mixed-initiative fact-checking.
\newblock In {\em Proceedings of the 31st annual {ACM} symposium on user interface software and technology}, pages 189--199.

\bibitem[Pennycook and Rand, 2022]{pennycook2022nudging}
Pennycook, G. and Rand, D.~G. (2022).
\newblock Nudging social media toward accuracy.
\newblock {\em {The Annals of the American Academy of Political and Social Science}}, 700(1):152--164.

\bibitem[Presser, 2025]{presser2025tiktok}
Presser, A. (2025).
\newblock Testing a new feature to enhance content on {TikTok}.
\newblock Accessed: 2025-09-24.

\bibitem[Quelle and Bovet, 2024]{quelle2024perils}
Quelle, D. and Bovet, A. (2024).
\newblock The perils and promises of fact-checking with large language models.
\newblock {\em {Frontiers in Artificial Intelligence}}, 7:1341697.

\bibitem[Saeed et~al., 2022]{saeed2022crowdsourced}
Saeed, M., Traub, N., Nicolas, M., Demartini, G., and Papotti, P. (2022).
\newblock Crowdsourced fact-checking at {Twitter}: how does the crowd compare with experts?
\newblock In {\em {Proceedings of the 31st {ACM} International Conference on Information \& Knowledge Management}}, pages 1736--1746.

\bibitem[Schmitt et~al., 2024]{schmitt2024role}
Schmitt, V., Villa-Arenas, L.-F., Feldhus, N., Meyer, J., Spang, R.~P., and M{\"o}ller, S. (2024).
\newblock The role of explainability in collaborative human-{AI} disinformation detection.
\newblock In {\em Proceedings of the 2024 {ACM} Conference on Fairness, Accountability, and Transparency}, pages 2157--2174.

\bibitem[Shaar et~al., 2020]{shaar2020known}
Shaar, S., Babulkov, N., Da~San~Martino, G., and Nakov, P. (2020).
\newblock That is a known lie: Detecting previously fact-checked claims.
\newblock In {\em {Proceedings of the 58th annual meeting of the Association for Computational Linguistics}}, pages 3607--3618.

\bibitem[Sharma et~al., 2023]{sharma2023towards}
Sharma, M., Tong, M., Korbak, T., Duvenaud, D., Askell, A., Bowman, S.~R., Cheng, N., Durmus, E., Hatfield-Dodds, Z., Johnston, S.~R., et~al. (2023).
\newblock Towards understanding sycophancy in language models.
\newblock {\em arXiv preprint arXiv:2310.13548}.

\bibitem[Shi et~al., 2019]{shi2019wisdom}
Shi, F., Teplitskiy, M., Duede, E., and Evans, J.~A. (2019).
\newblock The wisdom of polarized crowds.
\newblock {\em {Nature Human Behaviour}}, 3(4):329--336.

\bibitem[Straub et~al., 2023]{straub2023cost}
Straub, V.~J., Tsvetkova, M., and Yasseri, T. (2023).
\newblock The cost of coordination can exceed the benefit of collaboration in performing complex tasks.
\newblock {\em {Collective Intelligence}}, 2(2):26339137231156912.

\bibitem[Sunstein, 2001]{sunstein2001echo}
Sunstein, C.~R. (2001).
\newblock {\em Echo chambers: {Bush v. Gore}, impeachment, and beyond}.
\newblock {Princeton University Press Princeton, NJ}.

\bibitem[Tang et~al., 2024]{tang2024minicheck}
Tang, L., Laban, P., and Durrett, G. (2024).
\newblock {M}ini{C}heck: Efficient fact-checking of {LLM}s on grounding documents.
\newblock In Al-Onaizan, Y., Bansal, M., and Chen, Y.-N., editors, {\em {{Proceedings of the 2024 Conference on Empirical Methods in Natural Language Processing}}}, pages 8818--8847, Miami, Florida, USA. Association for Computational Linguistics.

\bibitem[Thorne et~al., 2018]{thorne2018fever}
Thorne, J., Vlachos, A., Christodoulopoulos, C., and Mittal, A. (2018).
\newblock {FEVER}: a large-scale dataset for fact extraction and verification.
\newblock {\em arXiv preprint arXiv:1803.05355}.

\bibitem[Truong et~al., 2025]{truong2025delayed}
Truong, B.~T., Kim, S., Nogara, G., Verdolotti, E., Sahneh, E.~S., Saurwein, F., Just, N., Luceri, L., Giordano, S., and Menczer, F. (2025).
\newblock Delayed takedown of illegal content on social media makes moderation ineffective.
\newblock {\em arXiv preprint arXiv:2502.08841}.

\bibitem[Tsvetkova et~al., 2024]{tsvetkova2024new}
Tsvetkova, M., Yasseri, T., Pescetelli, N., and Werner, T. (2024).
\newblock A new sociology of humans and machines.
\newblock {\em {Nature Human Behaviour}}, 8(10):1864--1876.

\bibitem[Vidgen and Derczynski, 2020]{vidgen2020directions}
Vidgen, B. and Derczynski, L. (2020).
\newblock Directions in abusive language training data, a systematic review: Garbage in, garbage out.
\newblock {\em {Plos one}}, 15(12):e0243300.

\bibitem[Vogels et~al., 2020]{americans2020Vogels}
Vogels, E., Perrin, A., and Anderson, M. (2020).
\newblock Most americans think social media sites censor political viewpoints.
\newblock Accessed: February 26, 2025.

\bibitem[Wang, 2017]{wang2017liar}
Wang, W.~Y. (2017).
\newblock "liar, liar pants on fire": A new benchmark dataset for fake news detection.
\newblock {\em arXiv preprint arXiv:1705.00648}.

\bibitem[Wojcik et~al., 2022]{wojcik2022birdwatch}
Wojcik, S., Hilgard, S., Judd, N., Mocanu, D., Ragain, S., Hunzaker, M., Coleman, K., and Baxter, J. (2022).
\newblock {Birdwatch}: Crowd wisdom and bridging algorithms can inform understanding and reduce the spread of misinformation.
\newblock {\em arXiv preprint arXiv:2210.15723}.

\bibitem[Woolley and Gupta, 2024]{woolley2024understanding}
Woolley, A.~W. and Gupta, P. (2024).
\newblock Understanding collective intelligence: Investigating the role of collective memory, attention, and reasoning processes.
\newblock {\em {Perspectives on Psychological Science}}, 19(2):344--354.

\bibitem[{World Health Organization}, 2020]{who_munich_security_conference}
{World Health Organization} (2020).
\newblock Director-general speeches: Munich security conference.
\newblock Accessed: 2026-04-07.

\bibitem[{X Corp.}, 2024]{communityNotes}
{X Corp.} (2024).
\newblock {X}.com {Community Notes Guide}.
\newblock Accessed: 2024-04-3.

\bibitem[{X Corp.}, 2026]{x2024communitynotesapi}
{X Corp.} (2026).
\newblock {Community Notes API Overview}.
\newblock Accessed: 2026-03-06.

\bibitem[Yang and Menczer, 2025]{yang2025accuracy}
Yang, K.-C. and Menczer, F. (2025).
\newblock Accuracy and political bias of news source credibility ratings by large language models.
\newblock In {\em Proceedings of the 17th {ACM} Web Science Conference 2025}, pages 127--137.

\bibitem[Yaniv, 2011]{yaniv2011group}
Yaniv, I. (2011).
\newblock Group diversity and decision quality: amplification and attenuation of the framing effect.
\newblock {\em {International Journal of Forecasting}}, 27(1):41--49.

\bibitem[Yasseri, 2025]{yasseri2025computational}
Yasseri, T. (2025).
\newblock Computational sociology of humans and machines: conflict and collaboration.
\newblock In {\em {Handbook of Computational Social Science}}, pages 809--822. {Edward Elgar Publishing Limited}.

\bibitem[Yasseri and Menczer, 2023]{yasseri2023can}
Yasseri, T. and Menczer, F. (2023).
\newblock Can crowdsourcing rescue the social marketplace of ideas?
\newblock {\em Communications of the {ACM}}, 66(9):42--45.

\bibitem[Yasseri et~al., 2012]{yasseri2012dynamics}
Yasseri, T., Sumi, R., Rung, A., Kornai, A., and Kert{\'e}sz, J. (2012).
\newblock Dynamics of conflicts in {Wikipedia}.
\newblock {\em {PloS one}}, 7(6):e38869.

\bibitem[Zhang et~al., 2025]{zhang2025commenotes}
Zhang, S., Wang, L., Shi, D., Chuai, Y., Chen, J., Chen, Y., Wang, Y., Wang, Y., Yi, X., and Li, H. (2025).
\newblock {Commenotes}: Synthesizing organic comments to support community-based fact-checking.
\newblock {\em arXiv preprint arXiv:2509.11052}.

\bibitem[Zhou et~al., 2024]{zhou2024correcting}
Zhou, X., Sharma, A., Zhang, A.~X., and Althoff, T. (2024).
\newblock Correcting misinformation on social media with a large language model.
\newblock {\em arXiv preprint arXiv:2403.11169}.

\end{thebibliography}

\clearpage
\renewcommand{\thetable}{S\arabic{table}}
\renewcommand{\thefigure}{S\arabic{figure}}

\setcounter{table}{0}  % Reset table counter for the appendix
\setcounter{figure}{0} 
\appendix

\section*{Supplementary Information for \\AI Feedback Enhances Community-Based Content Moderation through Engagement with Counterarguments\\ Saeedeh Mohammadi and Taha Yasseri\\}
\subsection*{Note Writing Experiment}
\begin{figure}[H]
  \centering
      \includegraphics[width=.8\textwidth]{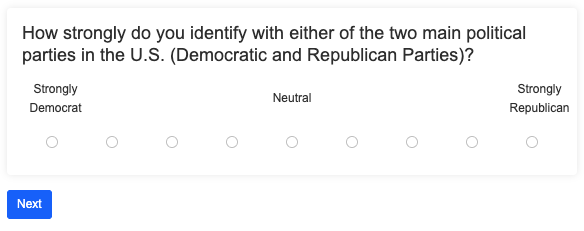}
\caption{A screenshot of the pre-experiment survey, where participants were asked to identify their partisanship on a nine-point scale. 
}
  \label{fig:note_writing_partisanship}
\end{figure}

In this experiment, participants first viewed an information sheet detailing the risks and benefits of participation, the study’s purpose, and information on ethics approval. After providing consent and entering their Prolific ID, participants were asked to indicate their political partisanship on a nine-point scale ranging from strongly Democrat to strongly Republican. A screenshot of this page is shown in Figure \ref{fig:note_writing_partisanship}. Participants who selected the neutral midpoint option were excluded from the experiment.

\begin{figure}[H]
  \centering
      \includegraphics[width=\textwidth]{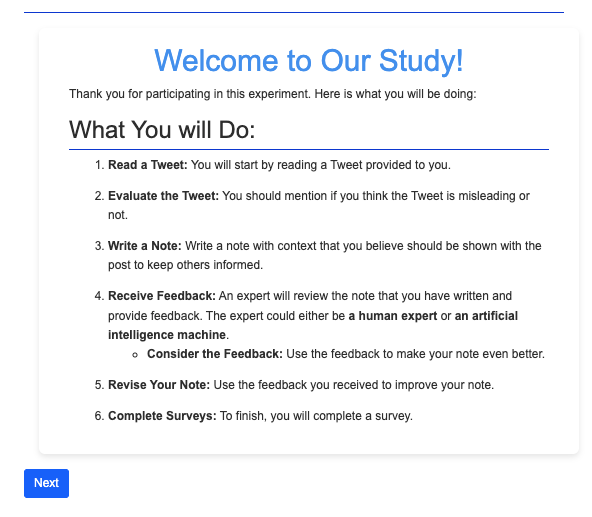}
\caption{A screenshot of the description of the study that participants read before the first task of the study.
}
  \label{fig:note_writing_description}
\end{figure}

Afterwards, participants were presented with a description of the study. At this stage, they were informed that the feedback they would receive could come from either a human expert or an artificial intelligence system. Figure \ref{fig:note_writing_description} shows a screenshot of this page.

\begin{figure}[H]
  \centering
      \includegraphics[width=\textwidth]{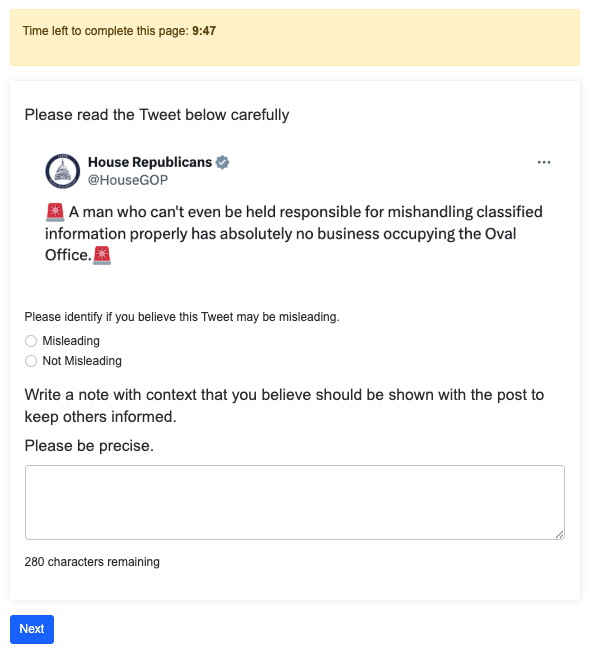}
\caption{A screenshot of the first task of the note writing experiment, which asks the participants to write a note to provide context to a social media post.
}
  \label{fig:note_writing_task1}
\end{figure}

In the first task of the experiment, participants were instructed to read a social media post and write a note to provide additional context and assess whether the post was misleading. The note was required to be under 280 characters. Participants were given 10 minutes to complete this task. Figure \ref{fig:note_writing_task1} shows an example of the task page.

\begin{figure}[H]
\centering
\includegraphics[width=\textwidth]{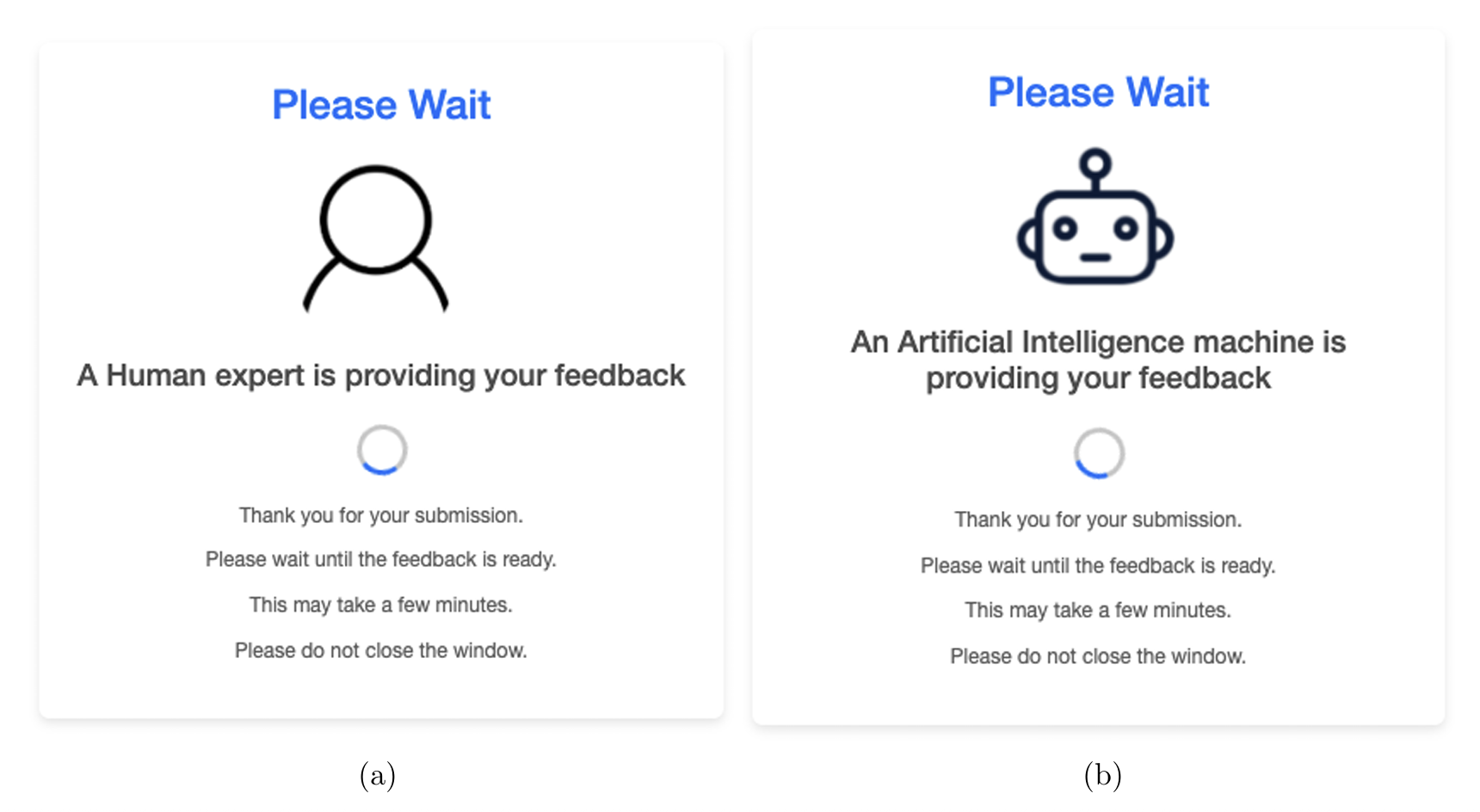}
\caption{The screenshot of the "please wait" page. (a) The feedback source is a human expert. (b) The feedback source is an artificial intelligence machine.
\label{fig:note_writing_please_wait}
}
\label{fig:test}
\end{figure}

After completing the first task, participants were shown a "please wait" page, which informed them about the source of the upcoming feedback. They remained on this page for 2.5 minutes to create the impression that the feedback might be provided by a human. Figure \ref{fig:note_writing_please_wait} displays two screenshots of this page, illustrating both feedback source conditions.

\begin{figure}[H]
  \centering
      \includegraphics[width=0.8\textwidth]{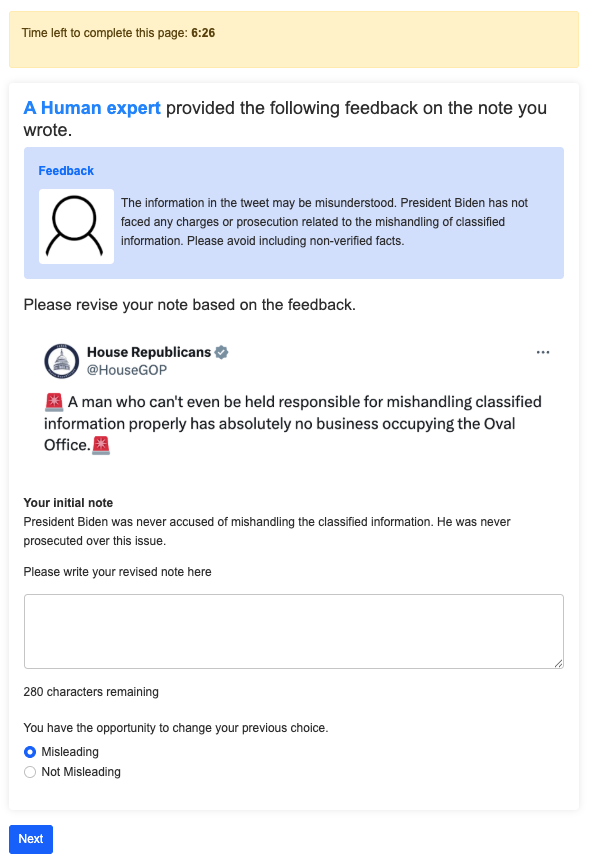}
\caption{A screenshot of the second task of the note writing experiment, where participants are given feedback on their notes and are asked to modify their notes, based on the feedback.
}
  \label{fig:note_writing_task2}
\end{figure}

Afterwards, participants were given seven minutes to review the feedback and revise their notes accordingly. During this stage, they were also allowed to change the post's classification (misleading or not misleading). Additionally, participants were reminded of the source of feedback. An example of this page is shown in Figure \ref{fig:note_writing_task2}.

\begin{figure}[H]
  \centering
      \includegraphics[width=0.8\textwidth]{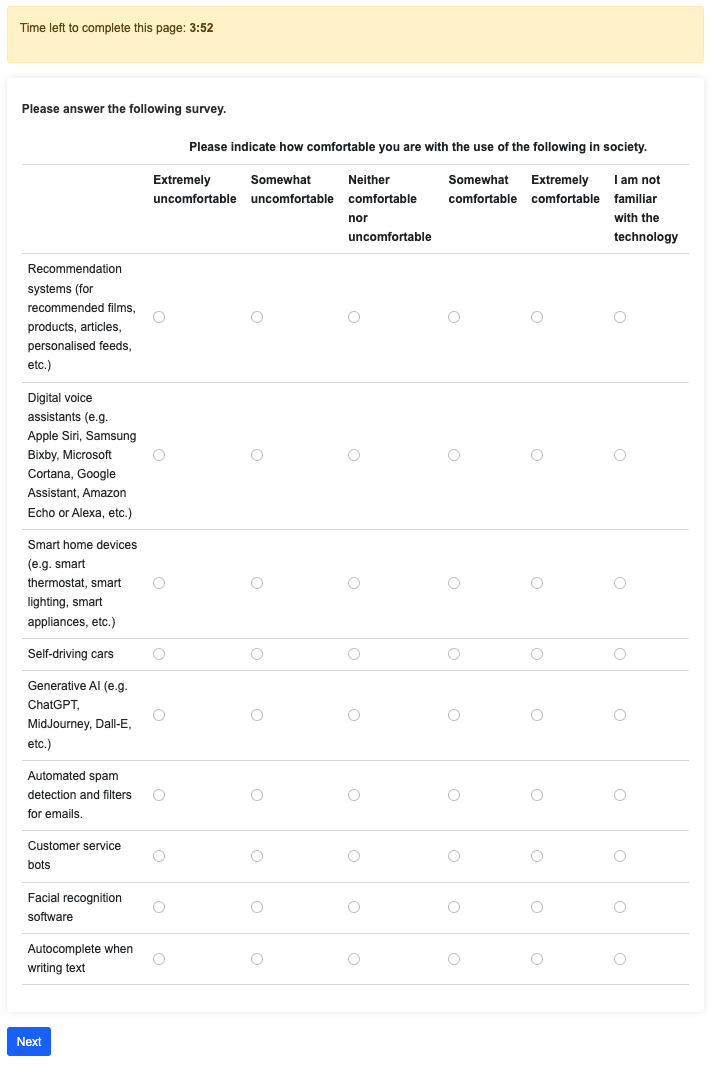}
\caption{A screenshot of the post-experiment survey
}
  \label{fig:note_writing_survey}
\end{figure}

After the second task, participants completed a post-experiment survey where they rated their comfort level with various technologies. Figure \ref{fig:note_writing_survey} shows the survey page displaying the full list of questions. Participants were given four minutes to complete the survey.

Subsequently, participants were asked to identify the source of the feedback as an attention check. Those who selected an incorrect option or chose “I do not remember” were excluded from the dataset. Finally, on the debrief page, participants were informed that all feedback was generated by an AI agent and that no human experts were involved.

\subsection*{Note Evaluation Experiment}
In the note evaluation experiment, participants were recruited to assess each note produced in the note writing experiment. Upon entering the study, participants first viewed an information sheet outlining the risks and benefits of participation, the study's purpose, and the ethics approval details. They then completed a consent form and provided their Prolific ID.

Afterwards, participants were asked to indicate their partisanship on a nine-point scale, consistent with the note-writing experiment. Those who selected the “neutral” option were excluded from the study.

\begin{figure}[H]
  \centering
      \includegraphics[width=0.8\textwidth]{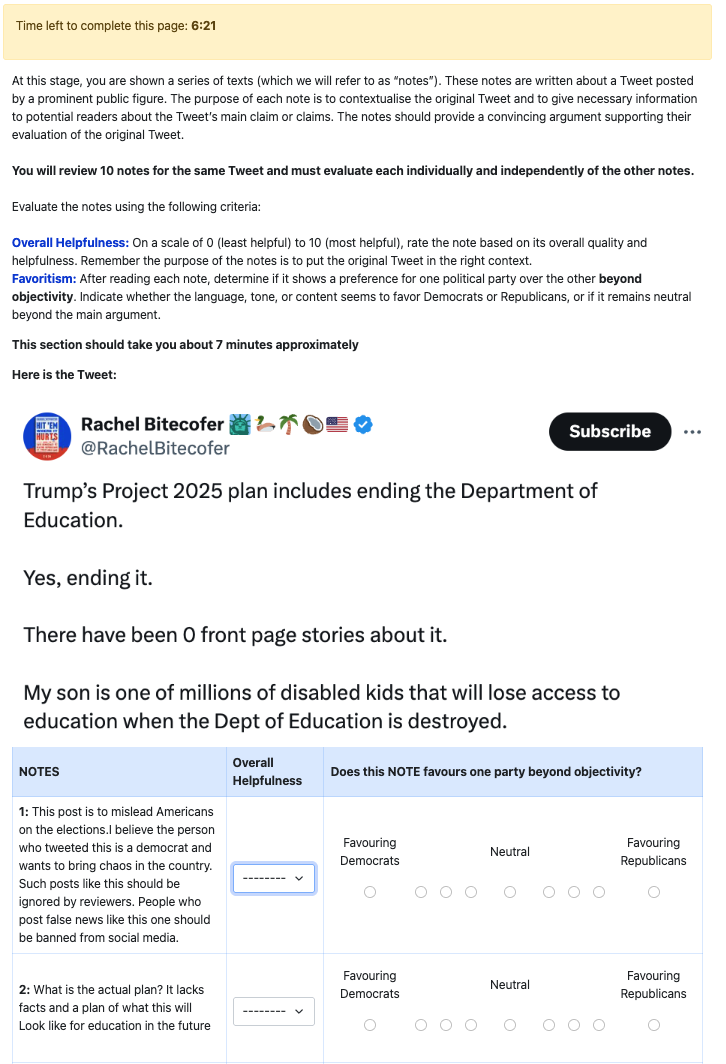}
\caption{A screenshot of the description provided to participants for evaluating the notes.
}
  \label{fig:note_evals}
\end{figure}

In the main task of the experiment, participants were shown a social media post accompanied by 10 notes about it. They were asked to rate each note on two dimensions: overall helpfulness and favouritism. Overall helpfulness was defined as the quality and usefulness of the note in providing appropriate context for the post, rated on a scale from 0 to 10. Favouritism was measured as the degree to which the note displayed bias favouring one political party over objectivity, on a nine-point scale ranging from strongly favouring Democrats to strongly favouring Republicans. Participants had 7 minutes to evaluate all 10 notes. Figure \ref{fig:note_evals} shows a screenshot of the task description. Each participant evaluated two different social media posts, each with its own set of 10 notes.

\subsection*{Protocol for selecting the Social Media Posts}
We selected 40 different posts from \url{X.com} for the note-writing experiment. These posts were drawn from the Community Notes dataset as of the end of August 2024. The selection followed the criteria below:

\begin{itemize}
\item The post contains at least one of the following keywords: Democrat, Republican, Biden, or Trump.
\item It has at least two notes tagged as "misinformed" in the Community Notes dataset.
\item The post is related to the United States.
\item The text is clear and easy to understand.
\item The post does not include a video or a link to an article required to understand its content.
\end{itemize}

Afterwards, the posts were classified as either pro-Democrat or pro-Republican based on their content. In total, we collected 20 pro-Democrat posts and 20 pro-Republican posts.

\subsection*{Feedback Prompts}
We used the OpenAI API (version 1.25.0) to prompt the GPT-4 model to generate feedback in the note-writing experiment. The \texttt{chat.completions} function was employed. The system message was set as follows:

\textit{``You are a helpful assistant who writes a note based on the instruction. The note should be limited to 280 characters. Do not use any hashtags or links in your response.''}

We set the temperature parameter to 1.0 to allow for variation in responses, while leaving all other parameters at their default values. Each participant received a single generated response without post-processing or regeneration.

The user prompt varied depending on the assigned feedback condition. The prompts were structured using two inputs: \texttt{\{tweet\}}, corresponding to the content of the social media post, and \texttt{\{comment\}}, corresponding to the participant’s note. We constructed the \texttt{\{comment\}} by appending the sentence ``The tweet is misleading'' or ``The tweet is not misleading'' to the participant’s original note, based on their classification.

The full prompts used in each condition are as follows:

\begin{itemize}
    \item \textbf{Argue:}  
    \textit{``Read the following tweet: \{tweet\}. Read the comment made in response to the tweet: \{comment\}. Argue with the comment while maintaining a consistent opposing stance.''}

    \item \textbf{Support:}  
    \textit{``Read the following tweet: \{tweet\}. Read the comment made in response to the tweet: \{comment\}. Support the comment's point. Help justify this comment and add additional arguments if possible.''}

    \item \textbf{Neutral:}  
    \textit{``Read the following tweet: \{tweet\}. Read the comment made in response to the tweet: \{comment\}. Rephrase the comment without adding new evidence or introducing contradictions.''}
\end{itemize}

The 280-character constraint was designed to mirror the format of Community Notes. Additionally, the model was instructed to avoid including links to reduce the risk of hallucinated or unverifiable content.

\subsection*{Participant Recruitment}

Participants for both experiments were recruited via the online research platform Prolific. We restricted the eligibility to residents of the United States. Participants were required to provide informed consent prior to participation.

\noindent
\textbf{Inclusion and Exclusion Criteria.}  
We included the participants if they (i) completed all required tasks, (ii) responded to all key measures, and (iii) demonstrated adequate understanding of the task instructions. We excluded participants who (i) did not provide consent, (ii) selected a neutral political affiliation, (iii) did not complete the tasks or were timed out, or (iv) failed the attention check.  

In the note-writing experiment, we additionally excluded cases in which participants did not revise their original note but instead responded directly to the feedback (e.g., writing “I agree” or “Correct” without modifying the note content), as this indicated non-compliance with the task instructions. This filter resulted in the removal of 28 cases.  

In the evaluation experiment, we further excluded participants who provided identical ratings across all items, as this indicated a lack of engagement with the task.

\noindent
\textbf{Sampling Design and Quotas.}  
We implemented a quota-based sampling design to ensure balanced partisan representation across conditions. In the note-writing experiment, each social media post was evaluated, within each treatment condition, by at least three self-identified Democrats and three self-identified Republicans. In the evaluation experiment, each note was rated by at least five self-identified Democrats and five self-identified Republicans.

Participants self-reported their political affiliation using a nine-point scale ranging from ``Strongly Democrat'' to ``Strongly Republican''. Only participants identifying with either partisan group were included in the final sample.

\noindent
\textbf{Sample Sizes.}  
The final sample for the note-writing experiment consisted of 893 participants (455 Democrats, 438 Republicans). The evaluation experiment included 1,354 participants (677 Democrats, 677 Republicans).

\noindent
\textbf{Compensation.}  
Participants were compensated at rates consistent with Prolific's fair-pay guidelines.

\subsection*{Statistical Modelling and Analytical Details}

We conducted the statistical analyses to examine (i) the determinants of engagement with AI-generated feedback and (ii) the effect of such engagement on changes in note quality.

\subsubsection*{Data Preprocessing}

The following preprocessing steps were applied to the ratings:

\begin{enumerate}
    \item \textbf{Standardisation of ratings:} Helpfulness scores were $z$-normalised at the rater--post level to control for individual rating tendencies.
    
    \item \textbf{Outlier handling:} For notes with at least five ratings per partisan group, the highest and lowest ratings were excluded prior to aggregation.
    
    \item \textbf{Aggregation:} Final helpfulness scores were computed as the mean of normalised ratings.
    
\end{enumerate}

\subsubsection*{Variable Construction and Coding Decisions}

We detail here the construction and coding of all variables used in the regression analyses.

\paragraph{Feedback Type.}
Feedback was categorized into three conditions: \textit{Argue}, \textit{Support}, and \textit{Neutral}. We operationalized this variable using indicator (dummy) variables, with \textit{Neutral} serving as the reference category in all regression models.

\paragraph{Participant Partisanship.}
Participant ideology was coded as a binary variable indicating whether a participant identified as a Democrat or a Republican. We used \textit{Republican} as the reference category.

\paragraph{Post Partisanship.}
Each post was classified according to its political leaning (Democrat or Republican). This variable was also coded as a binary variable, with \textit{Republican} as the reference category.

\paragraph{Participant--Post Alignment.}
To capture the relationship between the participant's ideology and the post's political leaning, we constructed a binary alignment variable based on their interaction:
\begin{itemize}
    \item \textbf{Co-partisan}: the participant and the post share the same political ideology
    \item \textbf{Cross-partisan}: the participant and the post have opposing political ideologies
\end{itemize}
This variable was derived from the interaction between participant partisanship and post partisanship. This was coded as a binary variable indicating whether a participant and post share political leanings. 

We used the interaction terms between feedback type and participant-post alignment in our statistical models, with the \textit{cross-partisan $\times$ neutral feedback} and the reference category.

\subsubsection*{Ordinal Logistic Regression Models}

To analyse changes in note quality, we estimated ordinal logistic regression models in which the dependent variable captures the direction of change in note helpfulness:

\begin{itemize}
    \item $I = -1$: Declined
    \item $I = 0$: Unchanged
    \item $I = 1$: Improved
\end{itemize}

Separate models were estimated for ratings from Democratic ($I_{\mathrm{D}}$) and Republican ($I_{\mathrm{R}}$) evaluators.

The key independent variable is the Feedback Acceptance rate ($FA$), treated as a continuous variable. Results are reported in Table~\ref{tab:logit_for_I_with_FA}.

To examine whether the effect of $FA$ varies across feedback types, we estimated models including interaction terms between $FA$ and feedback condition (Argue, Support, Neutral), with Neutral as the reference category. Results are reported in Table~\ref{tab:logit_I_FA_interaction_with_feedback}.

We estimated the models using the \texttt{polr()} function from the \texttt{MASS} package in R. We report log-odds, standard errors, and two-sided $p$-values based on Wald tests.

\subsubsection*{OLS Regression Models}

To analyse determinants of engagement with feedback, we estimated Ordinary Least Squares (OLS) regression models with Feedback Acceptance rate ($FA$) as the dependent variable.

\textbf{Predictors.}
\begin{itemize}
    \item Feedback type
    \item Participant partisanship
    \item Post type
    \item Alignment
    \item Interaction terms (e.g., Alignment $\times$ Feedback type) 
\end{itemize}

We estimated the models using the \texttt{lm()} function in R. We report coefficient estimates, standard errors, and two-sided $p$-values. The results are listed in Tables~S2, S3.

To assess the robustness of the ordinal logistic regression results, we conducted two sets of additional analyses.

First, we tested the proportional odds (parallel regression) assumption underlying the ordinal logistic models using the Brant test.

\noindent
\textbf{$FA$ as the sole predictor.}
For models based on Democratic ratings, the omnibus test does not indicate a violation of the proportional odds assumption ($\chi^2 = 6.86$, $df = 3$, $p = 0.08$). Consistent with this, none of the individual coefficients show statistically significant deviations ($p > 0.24$ for all terms), suggesting that the effect of Feedback Acceptance rate ($FA$) is stable across outcome thresholds.

Likewise, for models based on Republican ratings, the omnibus test does not indicate a violation of the proportional odds assumption ($\chi^2 = 3.68$, $df = 3$, $p = 0.30$). While the coefficient associated with the Neutral condition approaches significance ($p = 0.08$), it does not reach conventional thresholds, and all other terms are non-significant ($p > 0.14$). Overall, these results provide no evidence against the parallel regression assumption.

\noindent
\textbf{Feedback type and its interaction with $FA$ as predictors.}
For models based on Democratic ratings, the omnibus test indicates a violation of the proportional odds assumption ($\chi^2 = 11.7$, $df = 5$, $p = 0.04$). This violation is primarily driven by the Support condition ($p = 0.01$) and the interaction between Neutral feedback and Feedback Acceptance rate ($FA$) ($p = 0.02$), indicating that these effects may vary across outcome thresholds.

In contrast, for models based on Republican ratings, the proportional odds assumption is not violated (omnibus test: $\chi^2 = 3.75$, $df = 5$, $p = 0.59$), suggesting that the ordinal logistic specification is appropriate in this case.

To account for the violation observed in Democratic ratings, we additionally estimated multinomial logistic regression models that relax the proportional odds assumption. The results from these models are consistent with our main findings (Table~\ref{tab:multinom_ID}).

Second, we estimated several alternative ordinal logistic regression specifications to assess whether the main results are sensitive to model parameterisation. These models vary in whether they include the main effects of feedback type, the main effect of Feedback Acceptance rate ($FA$), and their interaction. Across specifications, the positive association between $FA$ and note improvement remains substantively similar, and the main interaction patterns are unchanged. Tables~S5, S6, S9, S10, S11, S12, and S13 report these alternative specifications.

\subsection*{Additional Figures}

\begin{figure}[H]
  \centering
      \includegraphics[width=\textwidth]{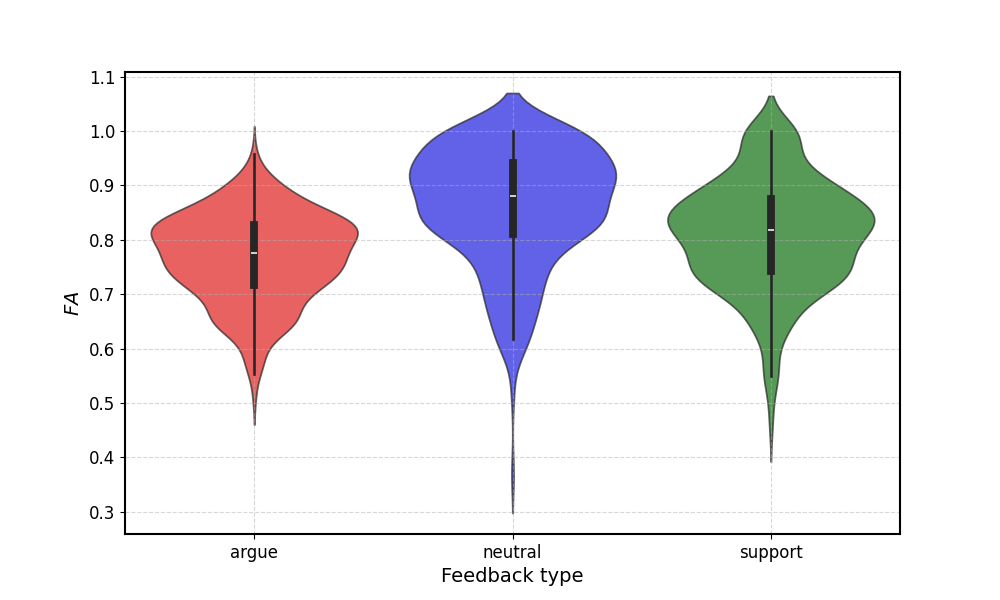}
\caption{The distribution of Feedback Acceptance rate ($FA$) grouped by the feedback type. The boxes inside the violin plots indicate the interquartile range, and the white line shows the median point.
}
  \label{fig:fbu_rate_by_treatments}
\end{figure}

\begin{figure}[H]
  \centering
      \includegraphics[width=\textwidth]{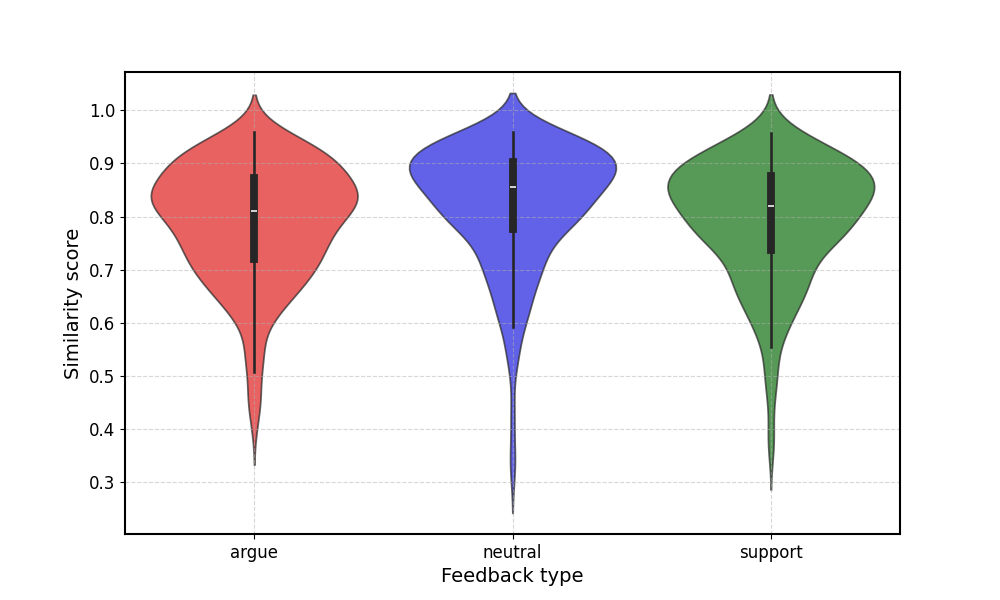}
\caption{The distribution of similarity score for initial notes and the notes written after the feedback, grouped by the feedback type. The boxes inside the violin plots indicate the interquartile range, and the white line shows the median point. 
}
  \label{fig:sim_score}
\end{figure}

\begin{figure}[H]
  \centering
      \includegraphics[width=\textwidth]{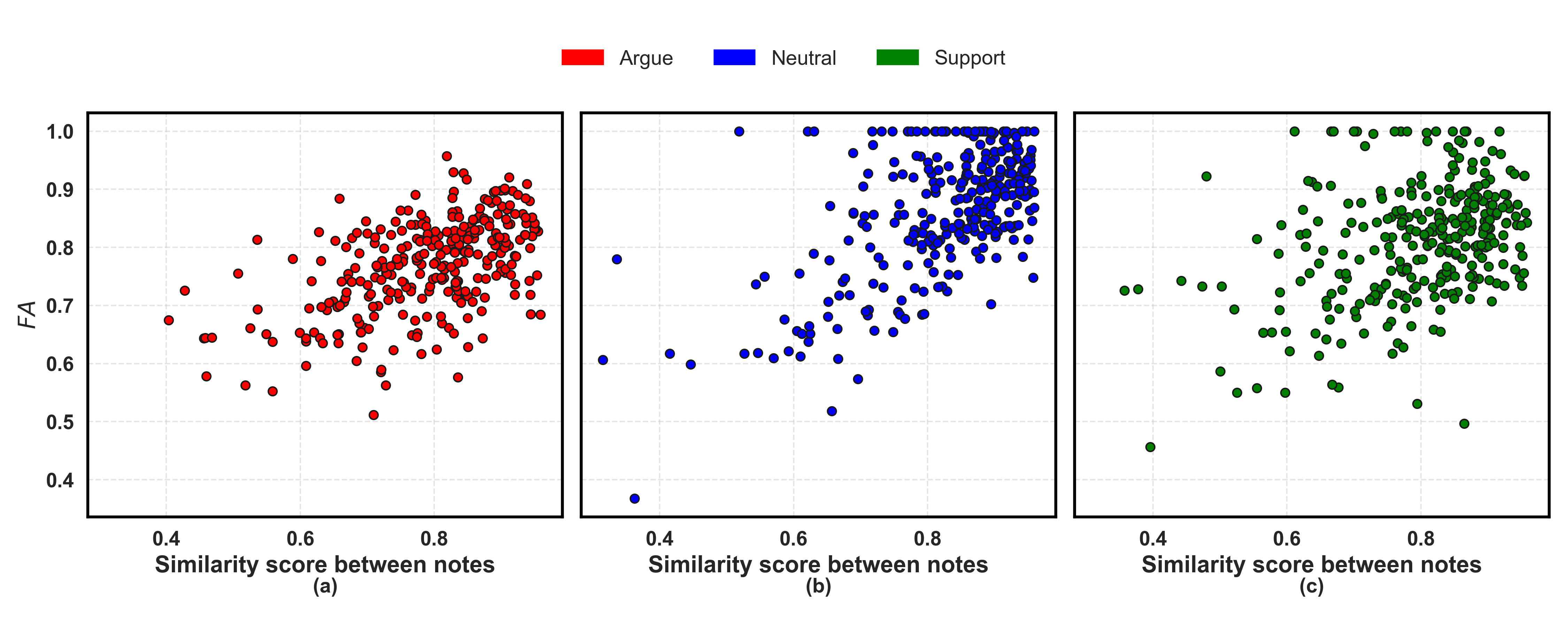}
\caption{Scatter plots of $FA$ versus similarity scores between initial and final notes across different feedback types. Panels (a), (b), and (c) correspond to argumentative, neutral, and supportive feedback conditions, respectively. The plots illustrate that a high similarity between two notes does not necessarily correspond to a high $FA$.}
  \label{fig:fa_vs_similarity}
\end{figure}

\begin{figure}[H]
  \centering
      \includegraphics[width=\textwidth]{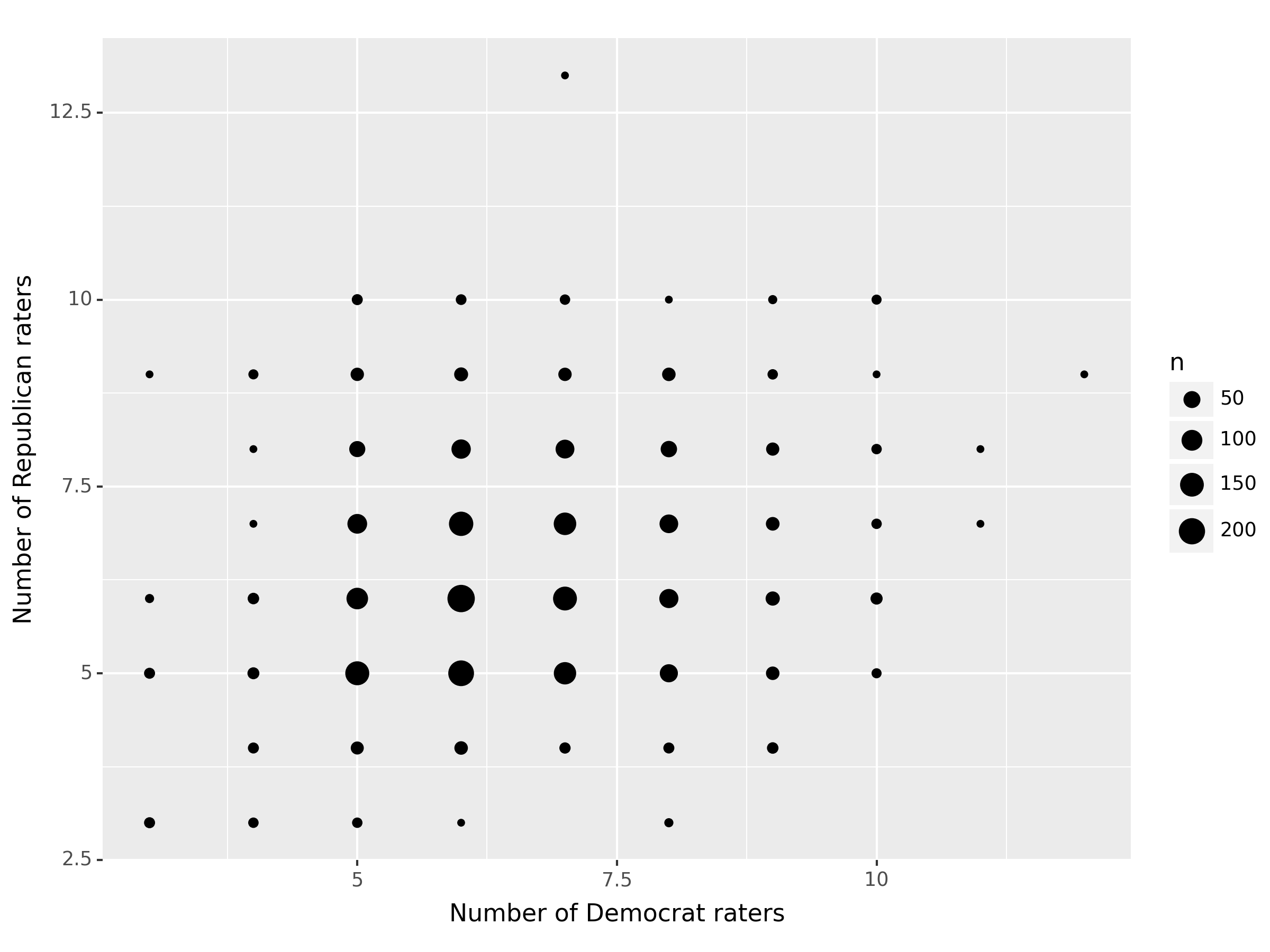}
\caption{Scatter plot showing the number of raters categorised by partisanship for each note in the dataset. The size of each node represents the volume of data at that point.
}
  \label{fig:eval_experiment}
\end{figure}

\subsection*{Additional Tables}
\begin{table}[H]
\centering
\caption{Results from ordinal logistic regression models predicting $I_\mathrm{D}$ and $I_\mathrm{R}$ using the Feedback Acceptance rate ($FA$) as the independent variable. Values in parentheses represent standard errors. Statistical significance is based on two-sided Wald tests.}

\label{tab:logit_for_I_with_FA}
\begin{tabular}{ccccc}
\hline
\multirow{2}{*}{Ordinal Logistical Regression} & \multicolumn{2}{c}{Dependent variable: $I_\mathrm{D}$} & \multicolumn{2}{c}{Dependent variable: $I_\mathrm{R}$} \\ \cline{2-5} 
                                               & Coefficient (log-odds)               & $P$                           & Coefficient (log-odds)              & $P$                            \\ \hline
$FA$                                           & 3.581(0.604)             & \textless{}0.001            & 2.498(0.599)              & \textless{}0.001             \\ \hline
Observations                                   & \multicolumn{2}{c}{893}                                & \multicolumn{2}{c}{893}                                \\ \hline
Pseudo $R^2$                                   & \multicolumn{2}{c}{0.019}                              & \multicolumn{2}{c}{0.009}                              \\ \hline
Deviance                                       & \multicolumn{2}{c}{1872.86}                            & \multicolumn{2}{c}{1924.63}                            \\ \hline
Null Deviance                                  & \multicolumn{2}{c}{1908.84}                            & \multicolumn{2}{c}{1942.18}                            \\ \hline
Log likelihood                                 & \multicolumn{2}{c}{-936.43}                            & \multicolumn{2}{c}{-962.32}                            \\ \hline
Akaike Inf. Crit.                              & \multicolumn{2}{c}{1878.86}                            & \multicolumn{2}{c}{1930.63}                            \\ \hline
\end{tabular}
\end{table}

\begin{table}[H]
\centering
\caption{ Results from Ordinary Least Squares (OLS) regression with Feedback Acceptance rate ($FA$) as the dependent variable. Independent variables include feedback type (Neutral as the reference), post type (Republican as the reference), and participant partisanship (Republican as the reference). Values in parentheses represent standard errors. Statistical significance is based on two-sided Wald tests.}

\label{tab:ols_FA_no_interaction}   
\begin{tabular}{ccc}
\hline
\multirow{2}{*}{Ordinary Least Squares (OLS) Regression} & \multicolumn{2}{c}{Dependent variable: $FA$} \\ \cline{2-3} 
                                                         & Coefficient          & $P$            \\ \hline
Feedback (Argue)                                         & -0.094(0.008)        & \textless{}0.001      \\ \hline
Feedback (Support)                                       & -0.052 (0.008)       & \textless{}0.001      \\ \hline
Post (D)                                                 & 0.001 (0.007)        & 0.880                 \\ \hline
Participant (D)                                          & 0.021 (0.007)        & 0.001                 \\ \hline
Constant                                                 & 0.850 (0.007)        & \textless{}0.001      \\ \hline
Observations                                             & \multicolumn{2}{c}{893}                      \\ \hline
$R^2$                                                    & \multicolumn{2}{c}{0.143}                    \\ \hline
Adjusted $R^2$                                           & \multicolumn{2}{c}{0.139}                    \\ \hline
Residual Std. Error                                      & \multicolumn{2}{c}{0.097 (df = 888)}         \\ \hline
F Statistics                                             & \multicolumn{2}{c}{36.899 (df = 4; 888)}     \\ \hline
\end{tabular}
\end{table}

\begin{table}[H]
\centering
\caption{ Ordinary Least Squares (OLS) regression results with Feedback Acceptance rate ($FA$) as the dependent variable. Predictors include feedback type (Neutral, Argue, Support) and the alignment between participant and post-partisanship (co- vs cross-partisan), as well as their interaction. The reference category is Cross-partisan × Neutral. Standard errors are shown in parentheses. Statistical significance is based on two-sided Wald tests.}
\label{tab:ols_FA_with_interaction}
\begin{tabular}{ccc}
\hline
\multirow{2}{*}{Ordinary Least Squares (OLS) Regression} & \multicolumn{2}{c}{Dependent variable $FA$} \\ \cline{2-3} 
                                                         & Coefficient         & $P$                     \\ \hline
Co-partisan × Neutral                                    & -0.019 (0.011)      & 0.090                 \\ \hline
Cross-partisan × Argue                                   & -0.102 (0.012)      & \textless{}0.001      \\ \hline
Co-partisan × Argue                                      & -0.104 (0.011)      & \textless{}0.001      \\ \hline
Cross-partisan × Support                                 & -0.059 (0.011)      & \textless{}0.001      \\ \hline
Co-partisan × Support                                    & -0.064 (0.011)      & \textless{}0.001      \\ \hline
Constant                                                 & 0.871 (0.008)       & \textless{}0.001      \\ \hline
Observations                                             & \multicolumn{2}{c}{893}                     \\ \hline
$R^2$                                                    & \multicolumn{2}{c}{0.135}                   \\ \hline
Adjusted $R^2$                                           & \multicolumn{2}{c}{0.130}                   \\ \hline
Residual Std. Error                                      & \multicolumn{2}{c}{0.098 (df = 887)}        \\ \hline
F Statistics                                             & \multicolumn{2}{c}{27.711 (df = 5; 887)}    \\ \hline
\end{tabular}
\end{table}

\begin{table}[H]
\centering
\caption{ Ordinal logistic regression predicting $I_\mathrm{D}$ and $I_\mathrm{R}$ using Feedback Acceptance rate ($FA$), feedback type (with \textit{Neutral} as the reference), and their interaction. Values in parentheses represent standard errors. Statistical significance levels are derived from two-sided Wald tests.
}
\label{tab:logit_I_FA_interaction_with_feedback}
\begin{tabular}{ccccc}
\hline
\multirow{2}{*}{Ordinal Logistical Regression} & \multicolumn{2}{c}{Dependent variable $I_\mathrm{D}$}                  & \multicolumn{2}{c}{Dependent variable $I_\mathrm{R}$} \\ \cline{2-5} 
                                               & Coefficient (log-odds)                         & $P$                               & Coefficient (log-odds)                     & $P$                  \\ \hline
Feedback (Argue)                               & -1.210 (1.362)                     & 0.374                             & -1.944 (1.346)                 & 0.148                \\ \hline
Feedback (Support)                             & \multicolumn{1}{l}{-0.583 (1.228)} & 0.635                             & -0.765 (1.238)                 & 0.536                \\ \hline
Feedback (Neutral) × $FA$                      & 2.643 (0.987)                      & 0.007                             & 1.534 (1.001)                  & 0.125                \\ \hline
Feedback (Argue) × $FA$                        & 3.713 (1.370)                      & 0.006                             & 3.817 (1.331)                  & 0.004                \\ \hline
Feedback (Support) × $FA$                      & 3.662 (1.086)                      & \textless{}0.001 & 2.237 (1.079)                  & 0.038                \\ \hline
Observations                                   & \multicolumn{2}{c}{893}                                                & \multicolumn{2}{c}{893}                               \\ \hline
Pseudo $R^2$                                   & \multicolumn{2}{c}{0.027}                                              & \multicolumn{2}{c}{0.011}                             \\ \hline
Deviance                                       & \multicolumn{2}{c}{1857.37}                                            & \multicolumn{2}{c}{1921.71}                           \\ \hline
Null Deviance                                  & \multicolumn{2}{c}{1908.84}                                            & \multicolumn{2}{c}{1942.18}                           \\ \hline
Log likelihood                                 & \multicolumn{2}{c}{-928.68}                                            & \multicolumn{2}{c}{-960.86}                           \\ \hline
Akaike Inf. Crit.                              & \multicolumn{2}{c}{1871.37}                                            & \multicolumn{2}{c}{1935.71}                           \\ \hline
\end{tabular}
\end{table}

\begin{table}[H]
\centering
\caption{Ordinal Logistic Regression Results for $I_\mathrm{D}$ and $I_\mathrm{R}$ with Interaction Terms Between Post Type (Democrat or Republican), Participant Partisanship (Democrat or Republican), and Feedback Type (Neutral, Argue, or Support). Post (R) × Participant (R) × Feedback (Neutral) is considered the reference. The values in the parentheses are the standard errors. Statistical significance levels (p-values) are derived from two-sided Wald tests.}
\label{tab:logit_for_I_interaction}
\resizebox{\textwidth}{!}{

\begin{tabular}{ccccc}
\hline
\multirow{2}{*}{Ordinal Logistical Regression}  & \multicolumn{2}{c}{Dependent variable $I_\mathrm{D}$} & \multicolumn{2}{c}{Dependent variable $I_\mathrm{R}$} \\ \cline{2-5} 
                                                & Coefficient (log-odds)                     & $P$                  & Coefficient (log-odds)                    & $P$                   \\ \hline
Post (D) × Participant (R) × Feedback (Neutral) & 0.449 (0.298)                  & 0.132                & 0.294 (0.307)                 & 0.338                 \\ \hline
Post (R) × Participant (D) × Feedback (Neutral) & 0.424 (0.299)                  & 0.156                & -0.202 (0.298)                & 4.9801                \\ \hline
Post (D) × Participant (D) × Feedback (Neutral) & 0.102 (0.307)                  & 0.739                & -0.318 (0.301)                & 0.291                 \\ \hline
Post (R) × Participant (R) × Feedback (Argue)   & -0.676 (0.304)                 & 0.026                & -0.292 (0.299)                & 0.328                 \\ \hline
Post (D) × Participant (R) × Feedback (Argue)   & -0.570 (0.313)                 & 0.068                & -0.317 (0.301)                & 0.293                 \\ \hline
Post (R) × Participant (D) × Feedback (Argue)   & -0.189 (0.306)                 & 0.537                & -0.403 (0.309)                & 0.192                 \\ \hline
Post (D) × Participant (D) × Feedback (Argue)   & -0.162 (0.295)                 & 0.583                & -0.569 (0.294)                & 0.0533                \\ \hline
Post (R) × Participant (R) × Feedback (Support) & 0.249 (0.302)                  & 0.409                & 0.240 (0.308)                 & 0.436                 \\ \hline
Post (D) × Participant (R) × Feedback (Support) & 0.144 (0.301)                  & 0.631                & -0.752 (0.297)                & 0.011                 \\ \hline
Post (R) × Participant (D) × Feedback (Support) & 0.644 (0.302)                  & 0.033                & -0.488 (0.298)                & 0.133                 \\ \hline
Post (D) × Participant (D) × Feedback (Support) & 0.277 (0.301)                  & 0.358                & -0.329 (0.297)                & 0.268                 \\ \hline
Observations                                    & \multicolumn{2}{c}{893}                               & \multicolumn{2}{c}{893}                               \\ \hline
Pseudo $R^2$                                    & \multicolumn{2}{c}{0.019}                             & \multicolumn{2}{c}{0.011}                             \\ \hline
Deviance                                        & \multicolumn{2}{c}{1872.58}                           & \multicolumn{2}{c}{1920.29}                           \\ \hline
Null Deviance                                   & \multicolumn{2}{c}{1908.84}                           & \multicolumn{2}{c}{1942.18}                           \\ \hline
Log likelihood                                  & \multicolumn{2}{c}{-936.29}                           & \multicolumn{2}{c}{-960.15}                           \\ \hline
Akaike Inf. Crit.                               & \multicolumn{2}{c}{1898.58}                           & \multicolumn{2}{c|}{1946.29}                          \\ \hline
\end{tabular}}

\end{table}

\begin{table}[H]
\centering
\caption{Ordinal Logistic Regression Results for $I_\mathrm{D}$ and $I_\mathrm{R}$ with dependent variables: Post Type (Democrat or Republican, with Republican as the reference category), Participant Partisanship (Democrat or Republican, with Republican as the reference category), and Feedback Type (Neutral, Argue, or Support, with Neutral as the reference category). The values in the parentheses are the standard errors. Statistical significance levels (p-values) are derived from two-sided Wald tests.}
\label{tab:logit_for_I_no_interaction}

\begin{tabular}{ccccc}
\hline
\multirow{2}{*}{Ordinal Logistical Regression} & \multicolumn{2}{c}{Dependent variable $I_\mathrm{D}$} & \multicolumn{2}{c}{Dependent variable $I_\mathrm{R}$} \\ \cline{2-5} 
                                               & Coefficient (log-odds)               & $P$                        & Coefficient (log-odds)                     & $P$                  \\ \hline
Post (D)                                       & -0.025 (0.125)           & 0.840                      & 0.145 (0.124)                  & 0.242                \\ \hline
Participant (D)                                & 0.251 (0.125)            & 0.045                      & -0.225 (0.124)                 & 0.069                \\ \hline
Feedback (Argue)                               & -0.644 (0.155)           & \textless{}0.001           & -0.332 (0.152)                 & 0.029                \\ \hline
Feedback (Support)                             & 0.077 (0.152)            & 0.612                      & -0.268 (0.151)                 & 0.076                \\ \hline
Observations                                   & \multicolumn{2}{c}{893}                               & \multicolumn{2}{c}{893}                               \\ \hline
Pseudo $R^2$                                   & \multicolumn{2}{c}{0.015}                             & \multicolumn{2}{c}{0.005}                             \\ \hline
Deviance                                       & \multicolumn{2}{c}{1879.29}                           & \multicolumn{2}{c}{1931.85}                           \\ \hline
Null Deviance                                  & \multicolumn{2}{c}{1908.84}                           & \multicolumn{2}{c}{1942.18}                           \\ \hline
Log likelihood                                 & \multicolumn{2}{c}{-939.65}                           & \multicolumn{2}{c}{-965.93}                           \\ \hline
Akaike Inf. Crit.                              & \multicolumn{2}{c}{1891.29}                           & \multicolumn{2}{c}{1943.85}                           \\ \hline
\end{tabular}
\end{table}

\begin{table}[H]
\centering
\caption{Ordinary Least Squares (OLS) Regression Results with $FA$ as the dependent variable and interaction terms for Feedback Type (Neutral, Argue, Support) and Co or Cross partisanship between the post and the participant (Cross-partisan × Neutral is the reference term) only for the data with the AI agent as the feedback source. The values in the parentheses are the standard errors. Statistical significance levels (p-values) are derived from two-sided Wald tests.}
\label{tab:ols_FA_with_interaction_AI}
\begin{tabular}{ccc}
\hline
\multirow{2}{*}{Ordinary Least Squares (OLS) Regression} & \multicolumn{2}{c}{Dependent variable $FA$}                            \\ \cline{2-3} 
                                                         & Coefficient                        & $P$                               \\ \hline
Co-partisan × Neutral                                    & -0.025 (0.017)                     & 0.133                             \\ \hline
Cross-partisan × Argue                                   & \multicolumn{1}{l}{-0.097 (0.017)} & \textless{}0.001 \\ \hline
Co-partisan × Argue                                      & -0.103 (0.017)                     & \textless{}0.001 \\ \hline
Cross-partisan × Support                                 & -0.051 (0.017)                     & 0.002                             \\ \hline
Co-partisan × Support                                    & -0.068 (0.017)                     & \textless{}0.001 \\ \hline
Constant                                                 & 0.873 (0.008)                      & \textless{}0.001 \\ \hline
Observations                                             & \multicolumn{2}{c}{444}                                                \\ \hline
$R^2$                                                    & \multicolumn{2}{c}{0.118}                                              \\ \hline
Adjusted $R^2$                                           & \multicolumn{2}{c}{0.108}                                              \\ \hline
Residual Std. Error                                      & \multicolumn{2}{c}{0.100 (df = 438)}                                   \\ \hline
F Statistics                                             & \multicolumn{2}{c}{11.674 (df = 5; 438)}                               \\ \hline
\end{tabular}
\end{table}

\begin{table}[H]
\centering
\caption{Ordinary Least Squares (OLS) Regression Results with $FA$ as the dependent variable and interaction terms for Feedback Type (Neutral, Argue, Support) and Co or Cross partisanship between the post and the participant (Cross-partisan × Neutral is the reference term) only for the data with the Human expert as the feedback source. The values in the parentheses are the standard errors. Statistical significance levels (p-values) are derived from two-sided Wald tests.}
\label{tab:ols_FA_with_interaction_Human}
\begin{tabular}{ccc}
\hline
\multirow{2}{*}{Ordinary Least Squares (OLS) Regression} & \multicolumn{2}{c}{Dependent variable $FA$} \\ \cline{2-3} 
                                                         & Coefficient         & $P$                   \\ \hline
Co-partisan × Neutral                                    & -0.014 (0.016)      & 0.381                 \\ \hline
Cross-partisan × Argue                                   & -0.107 (0.016)      & \textless{}0.001      \\ \hline
Co-partisan × Argue                                      & -0.106 (0.015)      & \textless{}0.001      \\ \hline
Cross-partisan × Support                                 & -0.066 (0.016)      & \textless{}0.001      \\ \hline
Co-partisan × Support                                    & -0.061 (0.016)      & \textless{}0.001      \\ \hline
Constant                                                 & 0.891 (0.011)       & \textless{}0.001      \\ \hline
Observations                                             & \multicolumn{2}{c}{449}                     \\ \hline
$R^2$                                                    & \multicolumn{2}{c}{0.157}                   \\ \hline
Adjusted $R^2$                                           & \multicolumn{2}{c}{0.147}                   \\ \hline
Residual Std. Error                                      & \multicolumn{2}{c}{0.096 (df = 443)}        \\ \hline
F Statistics                                             & \multicolumn{2}{c}{16.481 (df = 5; 443)}    \\ \hline
\end{tabular}
\end{table}

\begin{table}[H]
\centering
\caption{Multinomial logistic regression results predicting changes in note quality ($I_\mathrm{D}$) for Democratic evaluators. The reference category is $I = -1$ (declined). Coefficients are log-odds. Values in parentheses represent standard errors. $p$-values are based on two-sided Wald tests.}
\label{tab:multinom_ID}
\begin{tabular}{cccccc}
\hline
\multirow{2}{*}{Variable} & \multicolumn{2}{c}{$I = 0$ (Unchanged)} & \multicolumn{2}{c}{$I = 1$ (Improved)} \\ \cline{2-5}
 & Coefficient & $P$ & Coefficient & $P$ \\ \hline

Intercept & -2.997 (1.228) & 0.015 & -3.485 (1.302) & 0.007 \\\hline

Treatment (Argue) & 0.652 (1.792) & 0.716 & -0.355 (1.978) & 0.858 \\ \hline

Treatment (Support) & 3.631 (1.762) & 0.039 & 0.528 (1.881) & 0.779 \\ \hline

Neutral $\times FA$ & 4.765 (1.462) & 0.001 & 5.033 (1.542) & 0.001 \\ \hline

Argue $\times FA$ & 3.295 (1.704) & 0.053 & 4.730 (1.926) & 0.014 \\ \hline

Support $\times FA$ & 0.204 (1.583) & 0.898 & 4.591 (1.669) & 0.006 \\ \hline

Observations & \multicolumn{4}{c}{893} \\\hline
Pseudo $R^2$ (McFadden) & \multicolumn{4}{c}{0.034} \\ \hline
Log-Likelihood & \multicolumn{4}{c}{-922.35} \\ \hline
Akaike Inf. Crit.  & \multicolumn{4}{c}{1868.69} \\

\hline
\end{tabular}
\end{table}

\begin{table}[H]
\centering
\caption{Ordinal logistic regression on $I_\mathrm{D}$ and $I_\mathrm{R}$ with Feedback Acceptance rate ($FA$) and the interaction between feedback type and $FA$. The model is represented as $P(I_x \leq k \mid \text{$FA$}, \text{Feedback}, \text{Feedback} \times F_a)$, where $k \in \{-1, 0, 1\}$ and $x$ can be D (Democrats) or R (Republicans). The values in the parentheses are the standard errors. Statistical significance levels (p-values) are derived from two-sided Wald tests.}
\label{tab:logit_I_FA_interaction_with_feedback_model2}
\begin{tabular}{ccccc}
\hline
\multirow{2}{*}{Ordinal Logistical Regression} & \multicolumn{2}{c}{Dependent variable $I_\mathrm{D}$} & \multicolumn{2}{c}{Dependent variable $I_\mathrm{R}$} \\ \cline{2-5} 
                                               & Coefficient (log-odds)                     & $P$                  & Coefficient (log-odds)                     & $P$                  \\ \hline
$FA$                                           & 2.643 (0.987)                  & 0.007                & 1.534 (1.001)                  & 0.125                \\ \hline
Feedback (Argue)                               & -1.210 (1.362)                 & 0.374                & -1.944 (1.346)                 & 0.148                \\ \hline
Feedback (Support)                             & -0.583 (1.228)                 & 0.635                & -0.765 (1.238)                 & 0.537                \\ \hline
Feedback (Argue) × $FA$                        & 1.070 (1.684)                  & 0.525                & 2.282 (1.663)                  & 0.170                \\ \hline
Feedback (Support) × $FA$                      & 1.019 (1.463)                  & 0.486                & 0.703 (1.471)                  & 0.633                \\ \hline
Observations                                   & \multicolumn{2}{c}{893}                               & \multicolumn{2}{c}{893}                               \\ \hline
Pseudo $R^2$                                   & \multicolumn{2}{c}{0.027}                             & \multicolumn{2}{c}{0.011}                             \\ \hline
Deviance                                       & \multicolumn{2}{c}{1857.37}                           & \multicolumn{2}{c}{1921.71}                           \\ \hline
Null Deviance                                  & \multicolumn{2}{c}{1908.84}                           & \multicolumn{2}{c}{1942.18}                           \\ \hline
Log likelihood                                 & \multicolumn{2}{c}{-928.68}                           & \multicolumn{2}{c}{-960.86}                           \\ \hline
Akaike Inf. Crit.                              & \multicolumn{2}{c}{1871.37}                           & \multicolumn{2}{c}{1935.71}                           \\ \hline
\end{tabular}
\end{table}

\begin{table}[H]
\centering
\caption{Ordinal logistic regression on $I_\mathrm{D}$ and $I_\mathrm{R}$ with Feedback Acceptance rate ($FA$) and the interaction between feedback type and $FA$. The model is represented as $P(I_x \leq k \mid \text{$FA$}, \text{Feedback} \times F_a)$, where $k \in \{-1, 0, 1\}$ and $x$ can be D (Democrats) or R (Republicans). The values in the parentheses are the standard errors. Statistical significance levels (p-values) are derived from two-sided Wald tests.}
\label{tab:logit_I_FA_interaction_with_feedback_model3}

\begin{tabular}{ccccc}
\hline
\multirow{2}{*}{Ordinal Logistical Regression} & \multicolumn{2}{c}{Dependent variable $I_\mathrm{D}$} & \multicolumn{2}{c}{Dependent variable $I_\mathrm{R}$} \\ \cline{2-5} 
                                               & Coefficient (log-odds)               & $P$                        & Coefficient (log-odds)              & $P$                        \\ \hline
$FA$                                           & 3.233 (0.623)            & \textless{}0.001           & 2.443 (0.620)            & \textless{}0.001           \\ \hline
Feedback (Argue) × $FA$                        & -0.410 (0.204)           & 0.045                      & -0.104 (0.202)           & 0.607                      \\ \hline
Feedback (Support) × $FA$                      & 0.349 (0.187)            & 0.062                      & -0.167 (0.185)           & 0.366                      \\ \hline
Observations                                   & \multicolumn{2}{c}{893}                               & \multicolumn{2}{c}{893}                               \\ \hline
Pseudo $R^2$                                   & \multicolumn{2}{c}{0.027}                             & \multicolumn{2}{c}{0.009}                             \\ \hline
Deviance                                       & \multicolumn{2}{c}{1858.17}                           & \multicolumn{2}{c}{1923.81}                           \\ \hline
Null Deviance                                  & \multicolumn{2}{c}{1908.84}                           & \multicolumn{2}{c}{1942.18}                           \\ \hline
Log likelihood                                 & \multicolumn{2}{c}{-929.08}                           & \multicolumn{2}{c}{-961.9}                            \\ \hline
Akaike Inf. Crit.                              & \multicolumn{2}{c}{1868.17}                           & \multicolumn{2}{c}{1933.81}                           \\ \hline
\end{tabular}\end{table}

\begin{table}[H]
\centering
\caption{Ordinal logistic regression on $I_\mathrm{D}$ and $I_\mathrm{R}$ with the interaction between feedback type and $FA$. The model is represented as $P(I_x \leq k \mid \text{Feedback} \times F_a)$, where $k \in \{-1, 0, 1\}$ and $x$ can be D (Democrats) or R (Republicans). The values in the parentheses are the standard errors. Statistical significance levels (p-values) are derived from two-sided Wald tests.}
\label{tab:logit_I_FA_interaction_with_feedback_model4}
\begin{tabular}{ccccc}
\hline
\multirow{2}{*}{Ordinal Logistical Regression} & \multicolumn{2}{c}{Dependent variable $I_\mathrm{D}$} & \multicolumn{2}{c}{Dependent variable $I_\mathrm{R}$} \\ \cline{2-5} 
                                               & Coefficient (log-odds)               & $P$                        & Coefficient (log-odds)              & $P$                        \\ \hline
Feedback (Neutral) × $FA$                      & 3.233 (0.623)            & \textless{}0.001           & 2.443 (0.620)            & \textless{}0.001           \\ \hline
Feedback (Argue) × $FA$                        & 2.823 (0.701)            & \textless{}0.001           & 2.339(0.698)             & \textless{}0.001           \\ \hline
Feedback (Support) × $FA$                      & 3.582 (0.670)            & \textless{}0.001           & 2.276 (0.660)            & \textless{}0.001           \\ \hline
Observations                                   & \multicolumn{2}{c}{893}                               & \multicolumn{2}{c}{893}                               \\ \hline
Pseudo $R^2$                                   & \multicolumn{2}{c}{0.027}                             & \multicolumn{2}{c}{0.009}                             \\ \hline
Deviance                                       & \multicolumn{2}{c}{1858.17}                           & \multicolumn{2}{c}{1923.81}                           \\ \hline
Null Deviance                                  & \multicolumn{2}{c}{1908.84}                           & \multicolumn{2}{c}{1942.18}                           \\ \hline
Log likelihood                                 & \multicolumn{2}{c}{-929.08}                           & \multicolumn{2}{c}{-961.9}                            \\ \hline
Akaike Inf. Crit.                              & \multicolumn{2}{c}{1868.17}                           & \multicolumn{2}{c}{1933.81}                           \\ \hline
\end{tabular}

\end{table}

\begin{table}[H]
\centering
\caption{Ordinal logistic regression on $I_\mathrm{D}$ and $I_\mathrm{R}$ with Feedback Acceptance rate ($FA$) and $FA$. The model is represented as $P(I_x \leq k \mid \text{Feedback}, F_a)$, where $k \in \{-1, 0, 1\}$ and $x$ can be D (Democrats) or R (Republicans). The values in the parentheses are the standard errors. Statistical significance levels (p-values) are derived from two-sided Wald tests.}
\label{tab:logit_I_FA_interaction_with_feedback_model5}
\begin{tabular}{ccccc}
\hline
\multirow{2}{*}{Ordinal Logistical Regression} & \multicolumn{2}{c}{Dependent variable $I_\mathrm{D}$} & \multicolumn{2}{c}{Dependent variable $I_\mathrm{R}$} \\ \cline{2-5} 
                                               & Coefficient (log-odds)               & $P$                        & Coefficient (log-odds)              & $P$                        \\ \hline
$FA$                                           & 3.240 (0.648)            & \textless{}0.001           & 2.317 (0.644)            & \textless{}0.001           \\ \hline
Feedback (Argue) × $FA$                        & -0.331 (0.166)           & 0.046                      & -0.117 (0.164)           & 0.477                      \\ \hline
Feedback (Support) × $FA$                      & 0.272 (0.157)            & 0.083                      & -0.154 (0.155)           & 0.323                      \\ \hline
Observations                                   & \multicolumn{2}{c}{893}                               & \multicolumn{2}{c}{893}                               \\ \hline
Pseudo $R^2$                                   & \multicolumn{2}{c}{0.027}                             & \multicolumn{2}{c}{0.09}                              \\ \hline
Deviance                                       & \multicolumn{2}{c}{1858.01}                           & \multicolumn{2}{c}{1923.61}                           \\ \hline
Null Deviance                                  & \multicolumn{2}{c}{1908.84}                           & \multicolumn{2}{c}{1942.18}                           \\ \hline
Log likelihood                                 & \multicolumn{2}{c}{-929}                              & \multicolumn{2}{c}{-961.81}                           \\ \hline
Akaike Inf. Crit.                              & \multicolumn{2}{c}{1868.01}                           & \multicolumn{2}{c}{1933.61}                           \\ \hline
\end{tabular}\end{table}

%--/Paper--

\end{document}